\documentclass[superscriptaddress,preprintnumbers,nofootinbib,showpacs]{revtex4}

\usepackage{graphicx}
\usepackage{latexsym}
\usepackage[english]{babel}
\usepackage{amsmath}
\usepackage{amsthm}
\usepackage{epsfig}
\usepackage{color}
\usepackage{amssymb}

\begin{document}

\title{\Large Effect of the Trace Anomaly on the Cosmological Constant}

\preprint{ITP-UU-08/08, SPIN-08/08}


\author{Jurjen F. Koksma}
\email[]{J.F.Koksma@uu.nl} \affiliation{Institute for Theoretical
Physics (ITP) \& Spinoza Institute, Utrecht University, Postbus
80195, 3508 TD Utrecht, The Netherlands}

\author{Tomislav Prokopec}
\email[]{T.Prokopec@uu.nl} \affiliation{Institute for Theoretical
Physics (ITP) \& Spinoza Institute, Utrecht University, Postbus
80195, 3508 TD Utrecht, The Netherlands}

\begin{abstract}
It has been argued that the quantum (conformal) trace anomaly
could potentially provide us with a dynamical explanation of the
cosmological constant problem. In this paper, however, we show by
means of a semiclassical analysis that the trace anomaly does not
affect the cosmological constant. We construct the effective
action of the conformal anomaly for flat FLRW spacetimes
consisting of local quadratic geometric curvature invariants.
Counterterms are thus expected to influence the numerical value of
the coefficients in the trace anomaly and we must therefore allow
these parameters to vary. We calculate the evolution of the Hubble
parameter in quasi de Sitter spacetime, where we restrict our
Hubble parameter to vary slowly in time, and in FLRW spacetimes.
We show dynamically that a Universe consisting of matter with a
constant equation of state, a cosmological constant and the
quantum trace anomaly evolves either to the classical de Sitter
attractor or to a quantum trace anomaly driven one. When
considering the trace anomaly truncated to quasi de Sitter
spacetime, we find a region in parameter space where the quantum
attractor destabilises. When considering the exact expression of
the trace anomaly, a stability analysis shows that whenever the
trace anomaly driven attractor is stable, the classical de Sitter
attractor is unstable, and vice versa. Semiclassically, the trace
anomaly does not affect the classical late time de Sitter
attractor and hence it does not solve the cosmological constant
problem.
\end{abstract}

\maketitle

\section{Introduction}
\label{Introduction}

Recent observations have clearly indicated that the expansion of
the Universe is accelerating. According to Einstein's general
relativity, this can only be realised if the pressure of the
dominant component of the current Universe is negative. These
observations have triggered a renewed interest in the cosmological
constant problem (for recent reviews, see e.g.
\cite{Nobbenhuis:2004wn, Nobbenhuis:2006yf, Bousso:2007gp}). What
is usually referred to as the ``old'' cosmological constant
problem can be phrased as follows: why is the measured (effective)
cosmological constant extremely close to zero?

One approach dealing with the cosmological constant problem is
concerned with employing the effective field theory of gravity
\cite{Donoghue:1996ma, Donoghue:1997hx}. Lacking a full quantum
theory of gravity, an effective field theory of gravity adopts the
following point of view: in order to describe quantum phenomena at
very large and cosmologically relevant distances, the precise
physics at the shortest distance scales is irrelevant. In other
words, the effective field theory of gravity is the low energy
limit of quantum gravity. It combines classical general relativity
with knowledge of quantum field theory in curved spacetimes
\cite{Birrell:1982ix}.

In order to describe these long distance effects accurately, one
supplements the classical Einstein-Hilbert action with certain
additional contributions. One of these additions is the trace
anomaly or conformal anomaly which quantum field theories are
known to exhibit \cite{Capper:1974ic, Deser:1976yx, Brown:1976wc,
Dowker:1976zf, Tsao:1977tj, Duff:1977ay, Brown:1977pq,
Birrell:1982ix}. If the classical action is invariant under
conformal transformations of the metric, the resulting
stress-energy tensor is traceless. As an explicit example, one can
easily verify that the trace of a massless, conformally coupled
scalar field vanishes. In quantum field theory the stress-tensor
is promoted to an operator. A careful renormalisation procedure
renders its expectation value $\langle
T^{\mu}_{\phantom{\mu}\nu}\rangle$ finite. However, inevitably,
the renormalisation procedure results in general in a
non-vanishing trace of the renormalised stress-energy tensor.
Classical conformal invariance cannot be preserved at the quantum
level. Ever since its discovery, the trace anomaly has found many
applications in various areas in physics (see e.g.
\cite{Duff:1993wm}).

An alternative approach to the backreaction problem of quantum
fluctuations on the background spacetime deals with quantum fields
whose spectrum is nearly flat \cite{Ford:1984hs,
Antoniadis:1986sb, Tsamis:1992xa, Tsamis:1996qq, Tsamis:1996qk,
Abramo:1997hu, Finelli:2004bm, Bilandzic:2007nb, Janssen:2007ht,
Janssen:2008dw, Janssen:2008dp}. Consequently, the spectrum in the
infrared is not suppressed and is therefore expected to yield a
strong backreaction on the background spacetime. Examples of such
fields are the minimally coupled massless scalar and the graviton.

\subsection{The Connection between the Cosmological Constant and the Trace Anomaly}
\label{The Connection between the Cosmological Constant and the
Trace Anomaly}

Some authors stated that the trace anomaly could have effects on
dark energy and the cosmological constant problem
\cite{Bilic:2007gr, Schutzhold:2002pr}, whereas it has been argued
by other authors that the trace anomaly could potentially provide
us with a dynamical explanation of the cosmological constant
problem \cite{Tomboulis:1988gw, Antoniadis:1991fa,
Antoniadis:1992hz, Antoniadis:1998fi, Salehi:2000eu,
Antoniadis:2006wq}. Broadly speaking, the line of reasoning is as
follows (for a more in-depth review, we refer to
\cite{Antoniadis:2006wq}). The new, conformal degree of freedom is
usually parametrised by:
\begin{equation}\label{conformaltransformation}
g_{\mu\nu}(x)=e^{2\sigma(x)}\overline{g}_{\mu\nu}(x)\,.
\end{equation}
According to the authors of e.g. \cite{Antoniadis:2006wq}, the
trace anomaly cannot be generated from a local finite term in the
action, but rather stems from a non-local effective action that
generates the conformal anomaly by variation with respect to the
metric \cite{Riegert:1984kt}. It is this genuine non-locality of
the trace anomaly, revealing a large distance effect of quantum
physics, that is at the very foundation of its connection with the
effective field theory of gravity. One then argues that the new
conformal field should dynamically screen the cosmological
constant, thus solving the cosmological constant problem.

\subsection{The Semiclassical Approach to the Cosmological Constant and the Trace Anomaly}
\label{The Semiclassical Approach to the Cosmological Constant and
the Trace Anomaly}

The proposal advocated in \cite{Antoniadis:2006wq} is very
interesting and should be investigated further. Before studying
the effect of a new conformal degree of freedom
(\ref{conformaltransformation}), we feel that firstly a proper
complete analysis of the dynamics resulting from the effective
action of the trace anomaly should be performed. This is what we
pursue in this paper.

According to the Cosmological Principle the Universe is
homogeneous and isotropic on the largest and cosmologically
relevant scales. The CMB measurements \cite{Komatsu:2008hk}
constrain the inhomogeneities at order $10^{-4} \sim 10^{-5}$.
Moreover, the Universe appears to be spatially flat. Let us make
the following observations.

Firstly, the Cosmological Principle dictates the use of the
conformally flat FLRW metric $g_{\mu\nu}=
a^{2}(\eta)\eta_{\mu\nu}$. Hence, inhomogeneous fluctuations of
the metric tensor and in particular of the conformal part of the
metric tensor (\ref{conformaltransformation}) are observed and
expected to be small at the largest scales, comparable to the
Hubble radius (also in the early Universe).

Secondly, we are led to an essentially semiclassical analysis. The
vacuum expectation value of the stress energy tensor resulting in
the trace anomaly has been calculated semiclassically. In a
semiclassical analysis quantum fluctuations backreact on the
background spacetime. Phase transitions aside, quantum
fluctuations naturally affect the homogeneous background
\textit{homogenously}. It is a well-known fact that quantum
fluctuations can break certain symmetries present in de Sitter
\cite{Ford:1984hs, Antoniadis:1986sb, Tsamis:1992xa,
Tsamis:1996qq}, e.g. time translation invariance. However, we are
not aware of quantum fluctuations breaking the homogeneity and
isotropy of the background spacetime.

Finally, if quantum fluctuations compensate for or screen the
cosmological constant, we must have $T_{\mu\nu} \propto
g_{\mu\nu}$. Let us set: $T_{\mu\nu} = \theta(x) g_{\mu\nu}$.
Stress energy conservation and metric compatibility immediately
yield: $\nabla_{\mu} \theta(x)=\partial_{\mu} \theta(x)=0$. Hence
we conclude that $\theta(x)$ must be a constant: $\theta(x)=
\theta_{0}$. Only homogeneous vacuum fluctuations can compensate
the cosmological constant. Moreover, $T_{\mu\nu} = \theta_{0}
g_{\mu\nu}$ does not break any of the symmetries of a maximally
symmetric spacetime\footnote{Maximally symmetric spacetimes are de
Sitter, anti de Sitter and Minkowski spacetime.}. Hence, this form
cannot be used to study dynamical backreaction.

The arguments above motivate a semiclassical approach to examining
the connection between the cosmological constant and the trace
anomaly. Note that we do not consider a new, conformal degree of
freedom (\ref{conformaltransformation}). Hence, we do certainly
not exclude any possible effect this (inhomogeneous) conformal
degree of freedom might have on the cosmological constant.
However, it is plausible that in order to address the link between
the cosmological constant and the trace anomaly, a semiclassical
analysis suffices.

\subsection{The Modified Starobinsky Model}
\label{The Modified Starobinsky Model}

Another application of the conformal anomaly can be found in what
has become known as trace anomaly induced inflation: in the
absence of a cosmological constant, the trace anomaly could
provide us with an effective cosmological constant. Originally,
Starobinsky \cite{Starobinsky:1980te} realised that quantum one
loop contributions of massless fields can source a de Sitter
stage. Subsequently, the theory of trace anomaly induced inflation
received significant contributions from \cite{Hawking:2000bb,
Fabris:2000mq, Shapiro:2003gm, Shapiro:2004wt}. If one includes a
cosmological constant, the theory of anomaly induced inflation is
plagued by instabilities, which we will also come to address. The
Modified Starobinsky Model as advocated by \cite{Shapiro:2002nz,
Pelinson:2002ef, Pelinson:2003gn}, takes advantage of these
instabilities to account for a graceful exit from inflation. It is
argued that supersymmetry breaking changes the degrees of freedom
such that it destabilises the quantum anomaly driven attractor and
simultaneously stabilises the classical de Sitter attractor.

Improving on e.g. \cite{Pelinson:2002ef, Brevik:2006nh}, we
incorporate matter with a constant equation of state in the
Einstein field equations. We argue that it is simply inconsistent
not to include matter. Consider the following analogy: if we
examine an empty Universe with a cosmological constant only, there
is no dynamics and the (00) Einstein equation yields
$H^{2}=\Lambda/3$. Matter drives the dynamics and
$H^{2}\neq\Lambda/3$ can only be realised with
$\rho_{\mathrm{M}}\neq 0$.

In the literature, if one solves the trace of the Einstein field
equations in a Universe with a cosmological constant and trace
anomaly, one solves, however, in reality for the dynamics in a
Universe filled with radiation. For it is only radiation with
equation of state $w=1/3$ that does not contribute to the trace of
the Einstein field equation $T_{\mathrm{rad}}=0$. This point has
not been included in other papers. We will consider matter with
constant but otherwise arbitrary equation of state $w>-1$, and not
just (implicitly) radiation with $w=1/3$.

\subsection{Outline}
\label{Outline}

In this paper we show that the cosmological constant problem
cannot be solved by taking account of the trace anomaly alone. The
outline of this paper is as follows. In section \ref{Tracing the
Einstein field Equations and the Trace Anomaly} we recall the
basics of the conformal anomaly and discuss how to study its
effect on the evolution of the Universe by tracing the Einstein
field equations.

In section \ref{The Effective Action generating the Trace
Anomaly}, we derive the conformal anomaly from an effective action
in flat homogeneous FLRW spacetimes consisting of local quadratic
geometric curvature invariants. Since one usually adds infinite
counterterms to cancel the radiative one loop divergences, we do
not see any reason why we should exclude adding a Gauss-Bonnet
counterterm to cancel the anomaly in flat FLRW spacetimes. Even
though this term in the effective action is formally divergent, at
the level of the equation of motion it yields a finite result.
Hence, the coefficients multiplying the curvature invariants in
the trace anomaly are not uniquely specified by the anomaly. The
physical coefficient, i.e.: the parameter that can be measured,
receives contributions both from the trace anomaly and from
possible counterterms cancelling divergences from the underlying
(and yet unknown) fundamental theory. This motivates varying the
coupling parameters multiplying the curvature invariants in the
anomaly. We can thus study all possible effects of the anomaly on
the evolution of our Universe.

In section \ref{The Trace Anomaly in Quasi de Sitter spacetime} we
study the evolution of a quasi de Sitter Universe in the presence
of matter with constant equation of state, a cosmological constant
and the trace anomaly. In quasi de Sitter spacetime we assume,
loosely speaking, that the Hubble parameter is a slowly varying
function of time. Effectively, we truncate the expression of the
exact trace anomaly and discard higher order derivative
contributions.

In section \ref{The Trace Anomaly in FLRW Spacetimes} we
generalise our analysis and study the evolution of an FLRW
Universe again in the presence of matter with constant equation of
state, a cosmological constant and the trace anomaly. We examine
the exact trace anomaly and take all higher derivative
contributions into account. As the dimensionality of the phase
space increases, we must carefully perform a stability analysis of
the late time asymptotes.

\section{Tracing the Einstein field Equations and the Trace Anomaly}
\label{Tracing the Einstein field Equations and the Trace Anomaly}

\subsection{The Conformal Anomaly in Four Dimensions in FLRW
Spacetimes} \label{The Conformal Anomaly in Four Dimensions in
FLRW Spacetimes}

The trace anomaly or the conformal anomaly in four dimensions is
in general curved spacetimes given by \cite{Capper:1974ic,
Birrell:1982ix, Antoniadis:2006wq}:
\begin{equation}\label{TraceAnomaly1}
T_{\mathrm{Q}}\equiv \left\langle
T_{\phantom{\mu}\mu}^{\mu}\right\rangle = b F + b'
\left(E-\frac{2}{3}\Box R\right) + b'' \Box R \,,
\end{equation}
where:
\begin{subequations}
\label{TraceAnomaly2}
\begin{eqnarray}
E &\equiv& \mbox{}^{*}R_{\mu\nu\kappa\lambda}
\mbox{}^{*}R^{\mu\nu\kappa\lambda} = R_{\mu\nu\kappa\lambda}
R^{\mu\nu\kappa\lambda}- 4 R_{\mu\nu}
R^{\mu\nu} + R^{2} \label{TraceAnomaly2a} \\
F &\equiv& C_{\mu\nu\kappa\lambda} C^{\mu\nu\kappa\lambda} =
R_{\mu\nu\kappa\lambda} R^{\mu\nu\kappa\lambda}- 2 R_{\mu\nu}
R^{\mu\nu} + \frac{1}{3} R^{2} \label{TraceAnomaly2b} \,,
\end{eqnarray}
\end{subequations}
where as usual $R_{\mu\nu\kappa\lambda}$ is the Riemann curvature
tensor, $\mbox{}^{*}R_{\mu\nu\kappa\lambda}
=\varepsilon_{\mu\nu\alpha\beta}
R^{\alpha\beta}_{\phantom{\alpha\beta}\kappa\lambda}/2$ its dual,
$C_{\mu\nu\kappa\lambda}$ the Weyl tensor and $R_{\mu\nu}$ and $R$
the Ricci tensor and scalar, respectively. Note that $E$ is the
Gauss-Bonnet invariant. The general expression for the trace
anomaly can also contain additional contributions if the massless
conformal field is coupled to other long range gauge fields (see
e.g. \cite{Birrell:1982ix}). Finally, the parameters $b$, $b'$ and
$b''$ appearing in (\ref{TraceAnomaly1}) are dimensionless
quantities multiplied by $\hbar$ and are given by:
\begin{subequations}
\label{TraceAnomaly3}
\begin{eqnarray}
b &=& \frac{1}{120(4\pi)^{2}}\left( N_{S}+6N_{F}+ 12 N_{V}\right)
 \label{TraceAnomaly3a} \\
b' &=& -\frac{1}{360(4\pi)^{2}}\left( N_{S}+\frac{11}{2}N_{F}+ 62
N_{V}\right) \label{TraceAnomaly3b} \,,
\end{eqnarray}
\end{subequations}
where $N_{S}$, $N_{F}$ and $N_{V}$ denote the number of fields of
spin 0, 1/2 and 1 respectively ($\hbar=1$). It is important to
note that $b>0$ whereas $b'<0$ in general. It turns out that the
coefficient $b''$ is regularisation dependent and is therefore not
considered to be part of the true conformal anomaly. We take this
into account and study the effect of $b''$ on the stability of the
solutions we are about to derive. For definiteness, we will assume
that these parameters take their Standard Model values: $N_{S}=4$,
$N_{F}=45/2$ and $N_{V}=12$. Note if we were to include
right-handed neutrinos, $N_{F}=24$. One could also examine the
numerical value of the coefficients (\ref{TraceAnomaly3}) for the
late time Universe. Today's massless particle is just the photon,
hence $N_{V}=1$, $N_{S}=0$ and $N_{F}=0$.

Let us specialise to flat Friedmann-Lema\^itre-Robertson-Walker or
FLRW spacetimes in which the metric is given by $g_{\alpha\beta}=
\mathrm{diag} \left(-1,a^{2}(t),a^{2}(t),a^{2}(t)\right)$ where
$a(t)$ is the scale factor of the Universe in cosmic time $t$.
Recall that a conformal transformation leaves the Weyl tensor
invariant. Hence, in FLRW spacetimes $F=0$. Given the FLRW metric
one can easily verify that:
\begin{subequations}
\label{QuadraticCurvature}
\begin{eqnarray}
R^{2} &=& 36 \left[ \dot{H}^{2} +4(\dot{H}H^{2}+H^{4})\right]
 \label{QuadraticCurvaturea} \\
R_{\mu\nu} R^{\mu\nu} &=& 12 \left[ \dot{H}^{2}
+3(\dot{H}H^{2}+H^{4})\right]
 \label{QuadraticCurvatureb}\\
R_{\mu\nu\kappa\lambda} R^{\mu\nu\kappa\lambda} &=& 12 \left[
\dot{H}^{2} +2(\dot{H}H^{2}+H^{4})\right]
\label{QuadraticCurvaturec}\\
\Box R &=& - 6 \left[ \dddot{H} +7\ddot{H}H +4
\dot{H}^{2}+12\dot{H}H^{2}\right] \label{QuadraticCurvatured} \,.
\end{eqnarray}
\end{subequations}
Hence, the exact expression for the trace anomaly in FLRW
spacetimes in four dimensions reads:
\begin{equation}\label{TraceAnomaly4}
T_{\mathrm{Q}}= 4b' \left\{ \dddot{H} +7\ddot{H}H +4\dot{H}^{2} +
18\dot{H}H^{2} +6H^{4} \right\} -6b''\left\{ \dddot{H} +7\ddot{H}H
+4\dot{H}^{2} + 12\dot{H}H^{2}\right\}   \,.
\end{equation}
To capture the leading order dynamics we work in quasi de Sitter
spacetime and allow for a mildly time dependent Hubble parameter:
\begin{equation}\label{QdeSitterHubble}
\epsilon \equiv -\frac{\dot{H}}{H^{2}} = \mathrm{constant} \ll 1
\,,
\end{equation}
i.e.: we assume that $\epsilon$ is both small and time
independent. This would truncate the trace anomaly up to terms
linear in $\dot{H}$ yielding:
\begin{equation}\label{TraceAnomaly4b}
T_{\mathrm{Q}}= 24 b' \left\{ 3\dot{H}H^{2} +H^{4} \right\} -72b''
\dot{H}H^{2}  \,.
\end{equation}
We will examine both the exact form of the trace anomaly
(\ref{TraceAnomaly4}) and its truncated form
(\ref{TraceAnomaly4b}).

Truncating the expression for the trace anomaly is motivated by
the following realisation. In general backgrounds we need a
non-local effective action to generate the trace anomaly in the
equation of motion. The non-locality at the level of the effective
action corresponds to an expansion in derivatives at the level of
the equation of motion. Generally, higher derivative contributions
in an equation of motion have the tendency to destabilise a system
unless the initial conditions are highly fine-tuned. Formally,
this is known as the theorem of Ostrogradsky and its relevance to
Cosmology is outlined, for example, in \cite{Woodard:2006nt}.

Note that when $b''=2b'/3$ the truncated version is exact. We
discuss this further in section \ref{Case II: b''=2b'/3}. Finally,
note that although we have truncated equation
(\ref{TraceAnomaly4}) to obtain (\ref{TraceAnomaly4b}), equation
(\ref{TraceAnomaly4b}) is still covariant.

\subsection{The Dynamics driven by the Trace Anomaly}
\label{The Dynamics driven by the Trace Anomaly}

From the Einstein-Hilbert action:
\begin{equation}\label{EHaction}
S=S_{\mathrm{EH}}+S_{\mathrm{M}}=\frac{1}{16\pi G}\int d^{4}x
\sqrt{-g}\left(R-2\Lambda\right) + \int d^{4}x \sqrt{-g}
\mathcal{L}_{\mathrm{M}} \,,
\end{equation}
where:
\begin{equation}\label{EHaction2}
\mathcal{L}_{\mathrm{M}} = -\frac{1}{2}\partial_{\alpha}\phi(x)
\partial_{\beta} \phi(x) g^{\alpha\beta} - \frac{1}{2}m^{2}\phi^{2}(x) - V(\phi(x))\,,
\end{equation}
the Einstein field equations follow as usual as:
\begin{equation}\label{EFE1}
R_{\mu\nu}-\frac{1}{2}R g_{\mu\nu} +\Lambda g_{\mu\nu} = 8\pi G
T_{\mu\nu}\,,
\end{equation}
of which the trace can easily be verified to be:
\begin{equation}\label{EFETrace1}
R-4\Lambda = - 8\pi G T\,,
\end{equation}
where $T=T^{\mu}_{\phantom{\mu}\mu}$. If one considers an empty
Universe with a cosmological constant, the (00) Einstein field
equation acts as a constraint equation for the Hubble parameter
and one simply finds $H^{2}=\Lambda/3$ as usual. However, for a
non-empty Universe, the (00) Einstein field equation becomes a
dynamical constraint. The Bianchi identity for the left-hand side
of equation (\ref{EFE1}) straightforwardly results in
stress-energy conservation for the right-hand side:
\begin{equation}\label{Stressenergytensor1}
\nabla^{\mu}T_{\mu\nu}=0\,.
\end{equation}
Because of stress-energy conservation, the (00) and the (ij)
components of the Einstein field equations are not
independent\footnote{For example, stress-energy conservation
combined with the (00) Einstein field equation straightforwardly
yield the (ij) component of the Einstein field equations.}.
Therefore, any linear combination of the (00) and (ij) components
of the Einstein field equations combined with stress-energy
conservation suffice to describe the time evolution of the Hubble
parameter. In particular, the trace equation (\ref{EFETrace1}) and
stress-energy conservation (\ref{Stressenergytensor1}) contain all
relevant dynamics for $H$.

Let us set $\phi(x)=\phi_{\mathrm{cl}}(x)+\varphi(x)$ for the
quantum field in $S_{\mathrm{M}}$ and require that the classical
field obeys the equation of motion. Note that the quantum
perturbation $\varphi(x)$ does not obey this equation of motion.
We expand in terms of the quantum field and construct the
effective action as usual:
\begin{eqnarray}\label{effectiveaction1}
\exp\left[i \Gamma[\phi_{\mathrm{cl}}]\right] &=& \exp\left[i
S_{\mathrm{M}}[\phi_{\mathrm{cl}}]\right] \int \mathcal{D}\varphi
\exp\left[ i \left(\int_{x} \frac{\delta S_{\mathrm{M}}}{\delta
\phi_{\mathrm{cl}}(x)} \varphi(x) +\frac{1}{2} \int_{x,y}
\frac{\delta^{2} S_{\mathrm{M}}}{\delta
\phi_{\mathrm{cl}}(x)\phi_{\mathrm{cl}}(y) } \varphi(x)\varphi(y) +
\mathcal{O}(\varphi^{3}) \right)\right] \nonumber \\
&=& \exp\left[i S_{\mathrm{M}}[\phi_{\mathrm{cl}}]+ i
\Gamma_{\mathrm{Q}}[\phi_{\mathrm{cl}}]\right]\,.
\end{eqnarray}
The first contribution to the effective action corresponds to the
classical part of the action and $\Gamma_{\mathrm{Q}}
[\phi_{\mathrm{cl}}]$ is the contribution to the effective action
taking account of the vacuum fluctuations. The stress-energy
tensor now follows as:
\begin{equation}\label{Stressenergytensor2}
T_{\mu\nu}=-\frac{2}{\sqrt{-g}} \frac{\delta} {\delta g^{\mu\nu}}
\Gamma[\phi_{\mathrm{cl}}] = -\frac{2}{\sqrt{-g}} \frac{\delta}
{\delta g^{\mu\nu}} \left(S_{\mathrm{M}}[\phi_{\mathrm{cl}}] +
\Gamma_{\mathrm{Q}}[\phi_{\mathrm{cl}}]\right) \equiv
T_{\mu\nu}^{\mathrm{C}} + T_{\mu\nu}^{\mathrm{Q}}\,.
\end{equation}
Hence, there are both classical and quantum contributions to the
full stress-energy tensor. Classically, from the equation of
motion the scalar field obeys, we have:
\begin{equation}\label{Stressenergytensor3}
\nabla^{\mu}T_{\mu\nu}^{\mathrm{C}}=0\,.
\end{equation}
Hence, from (\ref{Stressenergytensor1}) we derive:
\begin{equation}\label{Stressenergytensor4}
\nabla^{\mu} T_{\mu\nu}^{\mathrm{Q}} =0\,.
\end{equation}
Concluding, due to stress-energy conservation at the classical
level and for the full stress-energy tensor, we have derived
stress-energy conservation for the quantum contributions as well.

Analogously to the classical stress-energy tensor, we can
symbolically write: $ T_{\phantom{\mu}\nu,\mathrm{Q}}^{\mu} =(
-\rho_{\mathrm{Q}}, p_{\mathrm{Q}},p_{\mathrm{Q}},
p_{\mathrm{Q}})$. Combining:
\begin{subequations}
\label{Stressenergytensor5}
\begin{eqnarray}
T_{\mathrm{Q}}(t) &=& -\rho_{\mathrm{Q}}(t) +3 p_{\mathrm{Q}}(t)
\label{Stressenergytensor5a}\\
\dot{\rho}_{\mathrm{Q}}(t) &=& - 3 H \left\{ \rho_{\mathrm{Q}}(t)
+ p_{\mathrm{Q}}(t)\right\} \label{Stressenergytensor5b} \,,
\end{eqnarray}
\end{subequations}
yields:
\begin{equation}\label{quantumdensity1}
\frac{d}{dt}\left[a^{4}(t)\rho_{\mathrm{Q}}(t)\right] = -
a^{4}(t)H(t) T_{\mathrm{Q}}(t)\,.
\end{equation}
We thus find (identical to \cite{Hawking:2000bb}):
\begin{subequations}
\label{Stressenergytensor6}
\begin{eqnarray}
\rho_{\mathrm{Q}}(t) &=& -\frac{1}{a^{4}(t)} \int^{t} d\tau
a^{4}(\tau) H(\tau) T_{\mathrm{Q}}(\tau)
\label{Stressenergytensor6a}\\
p_{\mathrm{Q}}(t) &=& \frac{1}{3}\left(T_{\mathrm{Q}}(t)+
\rho_{\mathrm{Q}}(t) \right) \label{Stressenergytensor6b} \,.
\end{eqnarray}
\end{subequations}
Although in general spacetimes it is not possible to perform this
integral, in cosmologically relevant FLRW spacetimes we can
\cite{Fischetti:1979ue}. If we consider the exact form of the
trace anomaly (\ref{TraceAnomaly4}) in flat FLRW spacetimes, we
can easily see that $\rho_{\mathrm{Q}}(t)$ should be of the
following form:
\begin{equation}\label{quantumdensity2}
\rho_{\mathrm{Q}}(t) = b'\left[ c_{1} \ddot{H}H + c_{2}
\dot{H}^{2}+c_{3} \dot{H}H^{2} +c_{4} H^{4}\right] + b''\left[
 c_{5} \ddot{H}H +c_{6}
\dot{H}^{2}+c_{7} \dot{H}H^{2} \right] \,.
\end{equation}
Upon inserting this ansatz into equation (\ref{quantumdensity1})
and equating the contributions at each order, we immediately find:
\begin{equation}\label{quantumdensity3}
\rho_{\mathrm{Q}}= 2 b'\left[
-2\ddot{H}H+\dot{H}^{2}-6\dot{H}H^{2} -3H^{4}\right] +3b''\left[
2\ddot{H}H -\dot{H}^{2}+6\dot{H}H^{2}\right]\,.
\end{equation}
Because $\rho_{\mathrm{Q}}$ can be expressed in a local form, we
would like to point out that this yields a local expression for
the stress-energy tensor too. Although quantum fluctuations of the
conformal field may still act non-locally on the background
spacetime, at the classical level the trace anomaly affects the
spacetime only locally. Again, when working in quasi de Sitter
spacetime where $\epsilon$ is both small and time independent, we
can truncate this expression for the quantum density finding:
\begin{equation}\label{quantumdensity4}
\rho_{\mathrm{Q}}= -6 b'\left[ 2\dot{H}H^{2} +H^{4}\right] +18 b''
\dot{H}H^{2}\,.
\end{equation}
Again, when $b''=2 b'/3$, this analysis becomes exact.

In the next sections, we solve the Einstein field equations in
FLRW spacetimes for a Universe in the presence of a) a non-zero
cosmological constant, b) the trace anomaly as a contribution to
the quantum stress-energy tensor and c) matter with constant
equation of state $\rho_{\mathrm{M}}=w p_{\mathrm{M}}$, where $w
> -1$. We thus consider non-tachyonic matter only, which does not
exclude \cite{Onemli:2004mb,Kahya:2006hc}, where the effect of a
finite period with $w < -1$ is investigated.

The trace anomaly enters the Einstein field equation naturally in
the trace equation (\ref{EFETrace1}). We can thus study the effect
of the trace anomaly on the evolution of a Universe with a
cosmological constant and matter. Hence, the trace equation of the
Einstein field equations is given by:
\begin{equation}\label{EFE2}
R-4\Lambda = - 8\pi G \left\{ T_{\mathrm{Q}} +
T_{\mathrm{M}}\right\}\,,
\end{equation}
where
\begin{equation}\label{EFE3}
T_{\mathrm{M}} = -\rho_{\mathrm{M}} + 3 p_{\mathrm{M}} =
\rho_{\mathrm{M}}\left(3\omega - 1\right)\,.
\end{equation}
Now, we can employ the (00) Einstein field equation from
(\ref{EFE1}) to express $\rho_{\mathrm{M}}$ in terms of
$\rho_{\mathrm{Q}}$ yielding:
\begin{equation}\label{EFEGeneral}
9(1+\omega) H^{2}(t)+6\dot{H}(t)-3(1+\omega)\Lambda = - 8\pi
G\left[T_{\mathrm{Q}}+ (1-3\omega)\rho_{\mathrm{Q}} \right]\,.
\end{equation}
The above equation governs the dynamics for the Hubble parameter
that we will solve in various interesting cases. For the anomalous
trace we can either take the exact expression
(\ref{TraceAnomaly4}) containing higher derivatives of the Hubble
parameter or its truncated version (\ref{TraceAnomaly4b}).
Likewise, for the quantum density we can either insert the full
expression (\ref{quantumdensity3}) or the truncated one
(\ref{quantumdensity4}). The reason for truncating the expression
for the quantum trace and density as outlined above is the
realisation that higher derivative contributions in an equation of
motion generally have the tendency to destabilise a system.

In the literature (see e.g. \cite{Pelinson:2003gn}), one only has
considered an empty Universe with a cosmological term and trace
anomaly. In reality, one solved for a radiation dominated
Universe, for only radiation ($\omega=1/3$) does not contribute to
the trace of the Einstein field equations. We incorporate matter
with constant equation of state parameter $\omega$.

Independently on whether one truncates the expressions for the
anomalous trace or quantum density, one can solve for the
asymptotes yielding the late time behaviour. Setting all time
derivatives of the Hubble parameter equal to zero, one can easily
solve for the two late time constants:
\begin{equation}\label{Asymptote1}
\left(H_{0}^{\mathrm{C,A}}\right)^2=\frac{-1 \pm \sqrt{1+64\pi G
b'\Lambda/3}}{32\pi G b'} \,.
\end{equation}
Here, $H_{0}^{\mathrm{C}}$ turns out to be the classical de Sitter
attractor, whereas $H_{0}^{\mathrm{A}}$ is a new, quantum anomaly
driven attractor. This result for the late time constants is
identical to the case where matter is absent
\cite{Pelinson:2002ef}. We can write the above expression in a
somewhat more convenient form by defining the dimensionless
parameter $\lambda$:
\begin{equation}\label{ScaleDependenceLambda}
\lambda=\frac{G\Lambda}{3} \,,
\end{equation}
that sets the scale for the cosmological constant $\Lambda$. In
the current epoch, $\lambda$ is extremely small, which allows us
to expand (\ref{Asymptote1}) finding\footnote{Note the
nomenclature in the literature is somewhat misleading. Rather than
calling (\ref{Asymptote2a}) the classical de Sitter attractor, it
would be more natural to denote it with the quantum corrected
classical attractor. Hence, the quantum attractor
(\ref{Asymptote2b}) should preferably be denoted by anomaly driven
attractor or Planck scale attractor. We will nevertheless adopt
the nomenclature existing in the literature.}:
\begin{subequations}\label{Asymptote2}
\begin{eqnarray}
H_{0}^{\mathrm{C}} &=& \sqrt{\frac{\Lambda}{3}}\left[1-8\pi
b'\lambda \right]
\label{Asymptote2a}\\
H_{0}^{\mathrm{A}} &=& \sqrt{\frac{-1}{16\pi G \, b'}
-\frac{\Lambda}{3}} \label{Asymptote2b}\,.
\end{eqnarray}
\end{subequations}
In the absence of a cosmological constant, the trace anomaly can
thus provide us with an inflationary scenario which has already
been appreciated by \cite{Starobinsky:1980te, Hawking:2000bb,
Shapiro:2003gm, Shapiro:2004wt}. Finally, note these asymptotes
are independent of $b''$.

\section{The Effective Action generating the Trace Anomaly}
\label{The Effective Action generating the Trace Anomaly}

The authors of \cite{Antoniadis:2006wq} are correct in saying that
the conformal anomaly cannot be generated by a finite effective
action built out of local quadratic geometric curvature invariants
only. We show that the trace anomaly can be generated by an
infinite effective action in flat FLRW spacetimes, that consists
only of local quadratic geometric curvature invariants. Although
infinite at the level of the effective action, we generate a
finite on shell contribution. We first write the trace anomaly in
conformal time $dt=a(\eta) d\eta$ such that the full conformal
anomaly (\ref{TraceAnomaly4}) reads:
\begin{equation}\label{TraceAnomaly5}
T_{\mathrm{Q}}= 24 b' \left(
\frac{a''}{a^{3}}\left(\frac{a'}{a^{2}}\right)^2 -
\left(\frac{a'}{a^{2}}\right)^4 \right) -6\left(b''-2b'/3 \right)
\left(\frac{a''''}{a^{5}}-4\frac{a'''}{a^4}\frac{a'}{a^2}-3\left(\frac{a''}{a^3}\right)^2
+6 \frac{a''}{a^{3}}\left(\frac{a'}{a^{2}}\right)^2 \right) \,.
\end{equation}
Here, dashes denote conformal time derivatives. In general, the
trace of a stress-energy tensor can be written as:
\begin{equation}\label{TraceStressTensor}
T = - \frac{2}{\sqrt{-g}} g^{\mu\nu} \frac{\delta S} {\delta
g^{\mu\nu}} = \frac{4}{a^{3}(\eta) V} \frac{\delta S} {\delta
a}\,.
\end{equation}
Here, $V$ is the (spatial) volume. The correct effective action
$\Gamma_{\mathrm{an}}$ that generates the trace anomaly in
spatially flat FLRW spacetimes is given by:
\begin{equation}\label{EffectiveAction1}
\Gamma_{\mathrm{an}} = \int d^{D}x \sqrt{-g}
\left\{\beta_{\mathrm{D}}E - 12\left(b''-2b'/3\right)
R^{2}\right\}\,,
\end{equation}
where
\begin{equation}\label{EffectiveAction2}
\beta_{\mathrm{D}} = b'\frac{1}{(D-4)}+\beta_{0}\,.
\end{equation}
Here $D$ is the dimension of the spacetime and $\beta_{0}$ is an
undetermined finite (and physically irrelevant) constant. Only
after variation we let $D \rightarrow 4$. The numerical factors in
(\ref{EffectiveAction1}) are chosen in accordance with the trace
of the Einstein field equations, i.e.: the trace of the variation:
\begin{equation}\label{Action1Loop}
g^{\mu\nu} \frac{\delta}{\delta g^{\mu\nu}} S=g^{\mu\nu}
\frac{\delta}{\delta g^{\mu\nu}}
\left[S_{\mathrm{EH}}+\Gamma_{\mathrm{an}}\right] = 0 \,,
\end{equation}
indeed yields:
\begin{equation}\label{Action1Loop2}
R-4\Lambda = - 8\pi G T\,.
\end{equation}
The effective action (\ref{EffectiveAction1}) merits some
clarifying remarks. Firstly, the variation of $R^{2}$
straightforwardly yields the $\Box R$ term in the conformal
anomaly. This term is not unique for one could have equally well
taken $R^{\mu\nu}R_{\mu\nu}$ into account\footnote{We can easily
split $R^{\mu\nu}R_{\mu\nu}$ in an $R^{2}$ contribution and a
Gauss-Bonnet term, that yields a vanishing surface term in four
dimensions.}. Secondly, note that the local effective action
depends solely on the scale factor. If one were to rewrite this
effective action covariantly in terms of the full metric, the
effective action would become non-local \cite{Deser:1976yx,
Riegert:1984kt}, also see \cite{Shapiro:2006sy, Shapiro:2008sf}.
It is well-known that the effective action of the conformal factor
can be written in a local form \cite{Davies:1977ti,
Fischetti:1979ue}. However, we have been able to rewrite this
expression in terms of the Gauss-Bonnet invariant multiplying an
infinite constant. Finally, note that the coefficient
(\ref{EffectiveAction2}) diverges in four dimensions. However, the
Gauss-Bonnet invariant in $D$ dimensions reads:
\begin{equation}\label{GaussBonnetDdims}
E = \left(
\frac{a'}{a^{2}}\right)^4\left[(D-8)(D-3)(D-2)(D-1)\right] +
\frac{a''}{a^{3}} \left( \frac{a'}{a^{2}}\right)^2\left[ 4
(D-3)(D-2)(D-1)\right] \,,
\end{equation}
which can be determined from the $D$ dimensional generalisations
of equation (\ref{QuadraticCurvature}). Note:
\begin{equation}\label{GaussBonnetDdims2}
-\frac{2}{\sqrt{-g}} g^{\mu\nu} \frac{\delta}{\delta
g^{\mu\nu}}\int d^{D}x \sqrt{-g}E = D(D-4)E \,.
\end{equation}
Clearly, only in four dimensions the Gauss-Bonnet term corresponds
to a total derivative that can be neglected. Hence, we choose the
divergent coefficient $\beta_{\mathrm{D}}$ such that the factor
$(D-4)$ cancels yielding a non-vanishing and finite contribution
at the level of the equation of motion when $D\rightarrow 4$. The
contribution from $\beta_{0}$ is identically zero when we let
$D\rightarrow 4$. Although divergent in the action, we have shown
that at the level of the equation of motion the term involving
$\beta_{\mathrm{D}}$ yields a finite contribution. All physical
measurements are performed on shell, hence we cannot exclude a
counterterm of the form $\beta_{\mathrm{D}}E$ in the effective
action.

This procedure can be debated. In the literature one usually
prefers a finite effective action. We abandon this assumption and
require a finite equation of motion only. We cannot think of a
physical measurement that distinguishes the two approaches and
therefore we argue that one should consider and examine all
possible effects the counterterms might have on the trace anomaly.

In homogeneous cosmology, $R^{2}$ and $E$ are the only local
quadratic geometric curvature invariants\footnote{Because $F=0$,
we can express $R_{\mu\nu\kappa\lambda} R^{\mu\nu\kappa\lambda}$
in terms of $R_{\mu\nu} R^{\mu\nu}$ and $R^{2}$. Hence, $R^{2}$
and $E$ are linearly independent. Note furthermore that if one
investigates inhomogeneities in the Universe, one measures
(statistical) correlation functions. They are translation
invariant as a consequence of the symmetry of the vacuum state and
therefore respect the symmetry of the background spacetime.}.
Divergences up to one loop in perturbative quantum gravity can be
cancelled only by counterterms of the form $R^{2}$ and $E$. The
physical coefficients multiplying $R^{2}$ and $E$ in the
renormalised one loop effective action receive contributions which
one can write as:
\begin{subequations}
\label{renormalisation1}
\begin{eqnarray}
\alpha_{\mathrm{phys}} R^{2} &=& \left[ \alpha_{\mathrm{anom}}(m) +
\alpha_{\mathrm{ct}}\right] R^{2}
 \label{renormalisation1a} \\
\beta_{\mathrm{phys}} E &=& \left[\beta_{\mathrm{anom}}(m) +
\beta_{\mathrm{ct}}\right] E\,,
\end{eqnarray}
\end{subequations}
where $\alpha_{\mathrm{anom}}(m)$ is the mass dependent finite
contribution from the trace anomaly, where $m=\left\{m_{i}
\right\}$ denotes the mass of the particles $i$. It is in
principle uniquely defined by the requirement that
$\alpha_{\mathrm{anom}}\rightarrow 0$ as $m\rightarrow \infty$.
Similarly, $\beta_{\mathrm{anom}}(m)$ is the (infinite) anomalous
contribution determined by the requirement that
$\beta_{\mathrm{anom}}\rightarrow 0$ as $m\rightarrow \infty$.
These requirements in principle fix $\beta_{\mathrm{anom}}(m)$ and
$\alpha_{\mathrm{anom}}(m)$ uniquely. While there is agreement in
the literature on the value of $\alpha_{\mathrm{anom}}(m)$ (also
used in this paper), disagreement exists on the value of
$\beta_{\mathrm{anom}}(m)$. Therefore we leave it unspecified. The
contributions $\alpha_{\mathrm{ct}}$ and $\beta_{\mathrm{ct}}$
correspond to the parts of the counterterms that remain when
eventual one loop divergences are cancelled\footnote{For a
calculation involving anomaly calculations around Minkowski
spacetime, we refer to \cite{Gorbar:2002pw, Gorbar:2003yt}. For
the calculation of the $\alpha_{\mathrm{anom}}(m)$ function for a
scalar field in de Sitter spacetime, we refer to
\cite{Birrell:1982ix}.}. It is however only the sum of these
terms, yielding $\alpha_{\mathrm{phys}}$ and
$\beta_{\mathrm{phys}}$, that is physical, i.e. measurable
\cite{Janssen:2008dw, Janssen:2008dp}. Since these parameters have
not been measured, we cannot simply assume that the coefficients
in the trace anomaly are just given by (\ref{TraceAnomaly3}). We
should allow these coefficients to vary in order to examine the
full effect of the trace anomaly on the evolution of our FLRW
Universe. This is precisely what we pursue in the following
sections.

\section{The Trace Anomaly in Quasi de Sitter spacetime}
\label{The Trace Anomaly in Quasi de Sitter spacetime}

In this section we work in quasi de Sitter spacetime, where we
treat $\epsilon$ as a small and time independent constant which
allows us to neglect higher order derivative contributions. The
reason for discarding higher order derivative contributions is
that they tend to destabilise a dynamical system, formally known
as the theorem of Ostrogradsky \cite{Woodard:2006nt}.

We have to distinguish two cases separately. In the spirit of
\cite{Antoniadis:2006wq}, the numerical value of the parameter
$b''$ occurring in the trace anomaly is not fixed because it is a
regularisation scheme dependent parameter. Generally, however, we
cannot exclude the presence of this term and we therefore allow it
to take different values. First, we allow for an unrestricted
value of $b''$ and secondly we set $b''=2 b'/3$. This case is
particularly interesting as this value of $b''$ sets the total
coefficient multiplying the $\Box R$ contribution in the trace
anomaly to zero.

\subsection{Case I: unrestricted value of $\mathbf{b''}$}
\label{Case I: unrestricted value of b''}

We thus insert the truncated expression for the trace anomaly
(\ref{TraceAnomaly4b}) and the quantum density
(\ref{quantumdensity4}) into the Einstein field equation
(\ref{EFEGeneral}). This yields:
\begin{equation}\label{EFETruncated}
9(1+\omega) H^{2}(t)+6\dot{H}(t)-3(1+\omega)\Lambda = - 8\pi
G\left[ \left\{ 12b'(5+3\omega)-54b''(1+\omega)
\right\}\dot{H}H^{2} +18b'(1+\omega)H^{4} \right]\,.
\end{equation}
This differential equation can be solved exactly. Separation of
variables yields:
\begin{equation}\label{EFE5}
t-t' = \frac{1}{(1+\omega)}\int_{H(t')}^{H(t)}dH \frac{2+ \alpha
H^{2}}{\Lambda-3H^{2} - 48\pi G b'H^{4}} \,,
\end{equation}
where $\alpha$ is conveniently defined as:
\begin{equation}\label{alphadef}
\alpha = 8\pi G\left\{ 4b'(5+3\omega)-18b''(1+\omega) \right\}\,.
\end{equation}
One can perform the integral in terms of logarithms where one has
to take the signs of the occurring parameters carefully into
account. The integral above gives:
\begin{eqnarray}\label{Hsolution1}
t-t' &=& \frac{1}{-48\pi G b'(1+\omega)\left\{
(H_{0}^{\mathrm{A}})^{2}-(H_{0}^{\mathrm{C}})^{2} \right\}}
\left[- \frac{1+\frac{1}{2}\alpha
(H_{0}^{\mathrm{A}})^{2}}{H_{0}^{\mathrm{A}}}
\left\{\log\left(\frac{H(t)+H_{0}^{\mathrm{A}}}{H(t)-
H_{0}^{\mathrm{A}}}\right) -
\log\left(\frac{H(t')+H_{0}^{\mathrm{A}}}{H(t')-
H_{0}^{\mathrm{A}}}\right)\right\} \right. \\
 && \qquad\qquad\qquad\qquad\qquad \qquad\qquad\quad
 + \left. \frac{1+\frac{1}{2}\alpha
(H_{0}^{\mathrm{C}})^{2}}{H_{0}^{\mathrm{C}}}
\left\{\log\left(\frac{H(t)+H_{0}^{\mathrm{C}}}{H(t)-
H_{0}^{\mathrm{C}}}\right) -
\log\left(\frac{H(t')+H_{0}^{\mathrm{C}}}{H(t')-
H_{0}^{\mathrm{C}}}\right)\right\} \right]\nonumber  \,.
\end{eqnarray}
Note the asymptotes of this analytic solution coincide with the
asymptotes obtained earlier in (\ref{Asymptote2}) as expected.

\begin{figure}
    \begin{minipage}[t]{.45\textwidth}
        \begin{center}
\includegraphics[width=\textwidth]{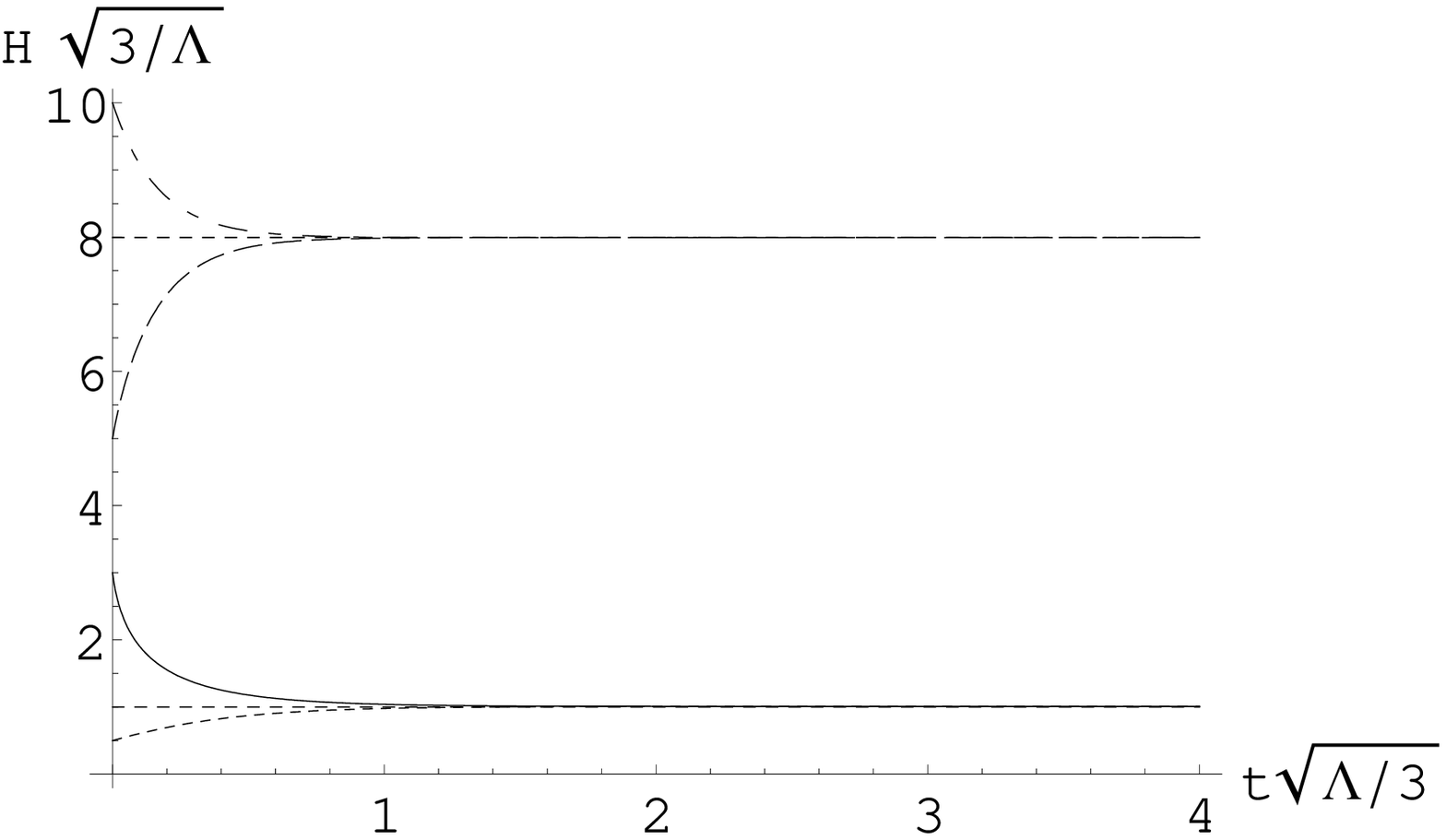}
   {\em \caption{Dynamics of the Hubble parameter in quasi de Sitter
   spacetime in the presence of a non-zero cosmological constant,
   the trace anomaly, and matter ($\omega=0$). Depending on the
   initial conditions, the Hubble parameter evolves to either the
   classical de Sitter or the quantum anomaly driven attractor. We
   have used $\lambda=1/50$, $b'=-0.015$ (Standard Model
   value) and $b''=0$. \label{fig:hubble1} }}
        \end{center}
   \end{minipage}
\hfill
    \begin{minipage}[t]{.45\textwidth}
        \begin{center}
\includegraphics[width=\textwidth]{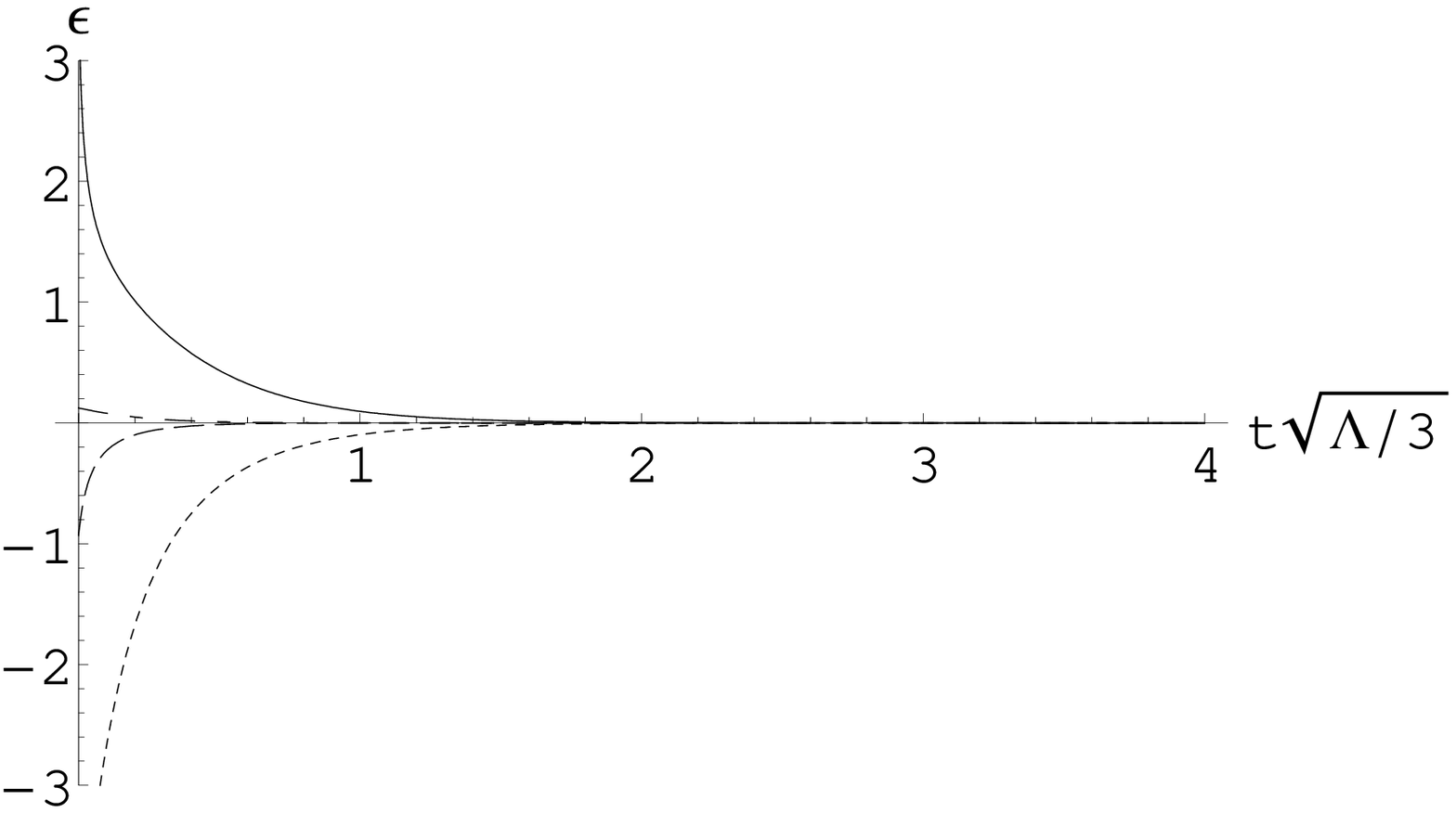}
   {\em \caption{Validity of the assumption $\epsilon\ll 1$. For the
   various initial conditions in figure \ref{fig:hubble1}, one can
   clearly see that this approximation is well justified. \label{fig:epsilon1} }}
        \end{center}
    \end{minipage}
\vskip 0.1cm
    \begin{minipage}[t]{.45\textwidth}
        \begin{center}
\includegraphics[width=\textwidth]{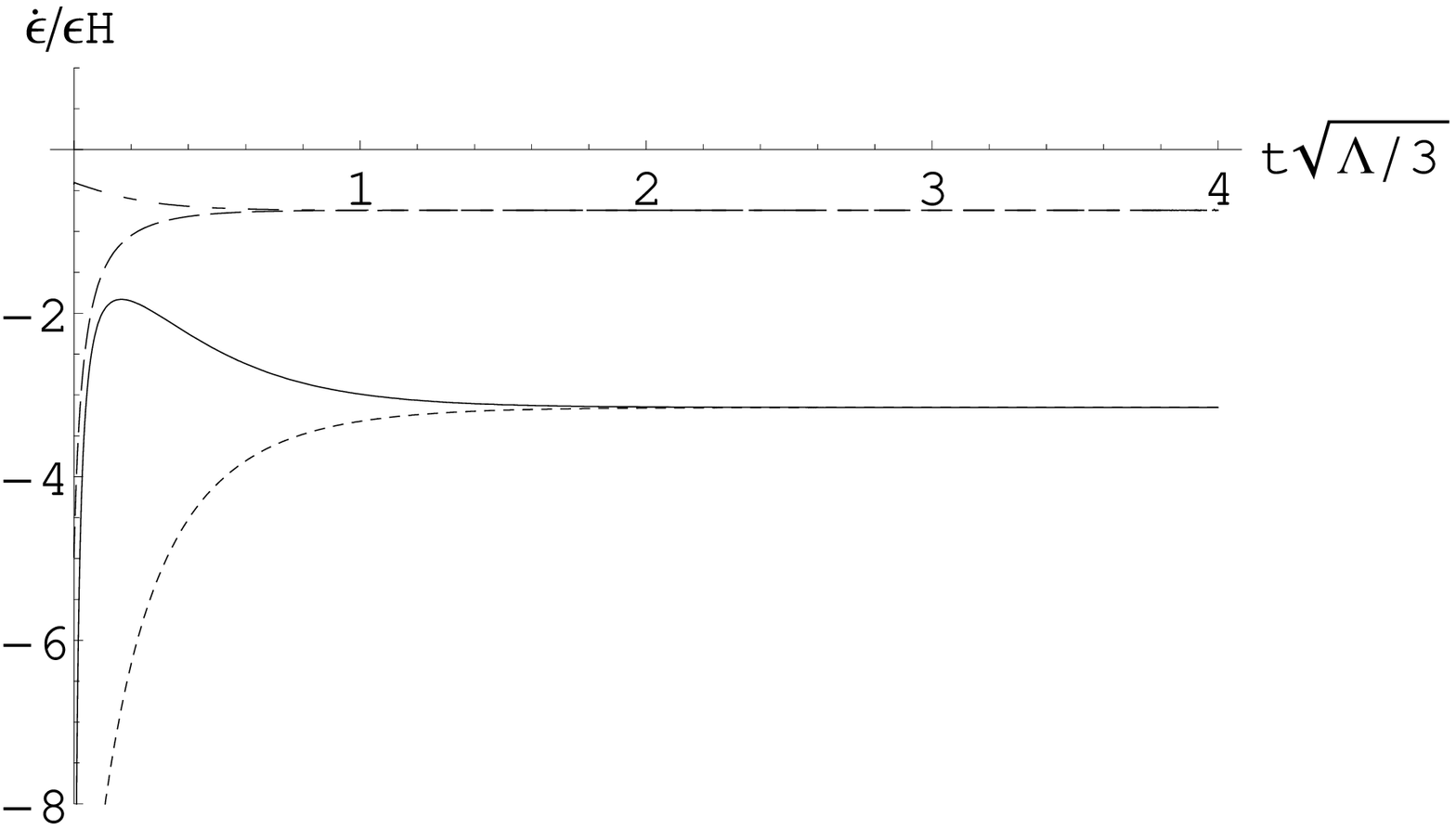}
   {\em \caption{Validity of the assumption $\dot{\epsilon}=0$. For the
   various initial conditions in figure \ref{fig:hubble1}, we have
   calculated $\dot{\epsilon}/(H \epsilon)$. Clearly, this condition is violated
   at all times. \label{fig:epsilondot1} }}
        \end{center}
    \end{minipage}
\hfill
    \begin{minipage}[t]{.45\textwidth}
        \begin{center}
\includegraphics[width=\textwidth]{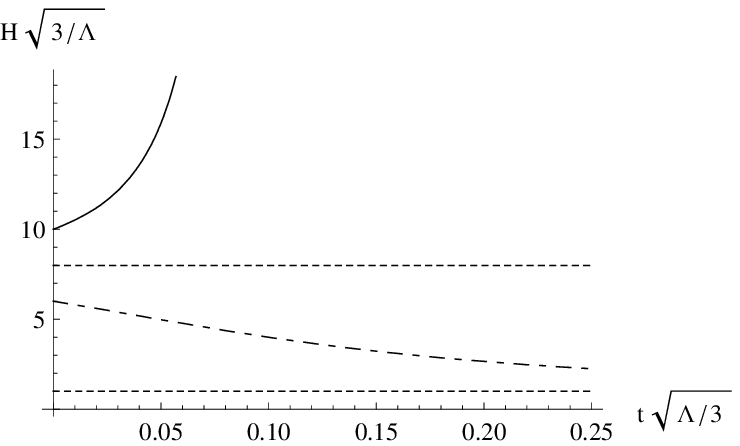}
   {\em \caption{Instability of the quantum anomaly driven attractor in quasi
   de Sitter spacetimes. We have used $\lambda=1/50$, $\omega=1/3$,
   $b''=7b'/6 < 5 b'/6$.
   \label{fig:QdSInstability} }}
        \end{center}
    \end{minipage}
\end{figure}

In figure \ref{fig:hubble1}, we numerically calculate the dynamics
of the Hubble parameter for various initial conditions. The two
asymptotes divide this graph into three distinct regions that are
not connected for finite time evolution. The region bounded by the
two asymptotes contains initial conditions for $H(t)$ such that
$H(t)$ grows for late times towards $H_{0}^{\mathrm{A}}$ and
initial conditions such that $H(t)$ asymptotes to the de Sitter
attractor $H_{0}^{\mathrm{C}}$. In figures \ref{fig:epsilon1} and
\ref{fig:epsilondot1}, we examine whether our approximation that
$\epsilon$ is both small and a constant is valid for calculating
the dynamics. In figure \ref{fig:epsilon1} one can clearly see
that $\epsilon\ll 1$ for late times. If $\dot{\epsilon}=0$, we
should have $\dot{\epsilon}/(H \epsilon)\ll 1$, an assumption that
is violated as depicted in figure \ref{fig:epsilondot1}. However,
already for a classical cosmological constant dominated Universe
with matter, a similar violation occurs.

Furthermore, note the existence of a branching point, an initial
condition for $H(t)$ such that for $H(0)>H_{\mathrm{BP}}$ the
Hubble parameter asymptotes to the quantum attractor
$H_{0}^{\mathrm{A}}$ and for $H(0)<H_{\mathrm{BP}}$ the Hubble
parameter decreases to the classical de Sitter attractor
$H_{0}^{\mathrm{C}}$. When rewriting equation of motion
(\ref{EFETruncated}) in terms of $\epsilon$ and noting that
$\epsilon$ should diverge exactly at the branching point, one
finds:
\begin{equation}\label{BranchingPoint}
H_{\mathrm{BP}}= \frac{1}{\sqrt{8\pi
G\left\{9(1+\omega)b''-2(5+3\omega)b'\right\}}} \,.
\end{equation}
We probe the dependence on scale by changing the numerical value
attached to $\lambda$ in equation (\ref{ScaleDependenceLambda}).
If we decrease $\lambda$, then also $\Lambda$ decreases which
results in a smaller $H_{0}^{\mathrm{C}}$. Also, it turns out that
both of the asymptotes are already reached much faster. This
improves the validity of the assumption $\epsilon\ll 1$, whereas
the assumption $\dot{\epsilon}/(H \epsilon)\ll 1$ is still
seriously violated at all times.

Let us study the analytical solution (\ref{Hsolution1}) more
closely. In particular, it is interesting to derive the high and
low energy limits of this solution. Of course, one naively expects
in the high or low energy limit to flow towards the quantum or
classical attractor, respectively. However, the analysis turns out
to be somewhat more subtle. We will show that under a certain
condition, the quantum anomaly driven attractor becomes unstable.
Although solution (\ref{Hsolution1}) looks complicated at first
glance, it simplifies when defining:
\begin{subequations}
\label{LimitDefs}
\begin{eqnarray}
\Omega &=& 3 (1+\omega)\left[1+ 32\pi b'\lambda\right]
\label{LimitDefs1} \\
A &=& \frac{1+\frac{1}{2}\alpha
(H_{0}^{\mathrm{A}})^{2}}{H_{0}^{\mathrm{A}}}
\label{LimitDefs2} \\
B &=& \frac{1+\frac{1}{2}\alpha
(H_{0}^{\mathrm{C}})^{2}}{H_{0}^{\mathrm{C}}} \label{LimitDefs3}
\,.
\end{eqnarray}
We define the initial conditions at $t'$ as:
\begin{eqnarray}
c_{1} &=& \log\left(\frac{H(t')+H_{0}^{\mathrm{A}}}{\left| H(t')-
H_{0}^{\mathrm{A}}\right|}\right)
\label{LimitDefs4} \\
c_{2} &=& \log\left(\frac{H(t')+H_{0}^{\mathrm{C}}}{\left| H(t')-
H_{0}^{\mathrm{C}}\right|}\right) \,. \label{LimitDefs5}
\end{eqnarray}
\end{subequations}
Note that $\Omega >1$, generically. With these definitions,
equation (\ref{Hsolution1}) reduces to:
\begin{equation}\label{Hsolution2}
\Omega \left(t - t'\right) = - A \left[
\log\left(\frac{H(t)+H_{0}^{\mathrm{A}}}{\left| H(t)-
H_{0}^{\mathrm{A}} \right| }\right) - c_{1} \right] + B \left[
\log\left(\frac{H(t)+H_{0}^{\mathrm{C}}}{\left| H(t)-
H_{0}^{\mathrm{C}}\right| }\right) - c_{2} \right] \,,
\end{equation}
In the high energy limit, we set:
\begin{equation}\label{Highlimit}
\delta(t) = \frac{H(t) - H_{0}^{\mathrm{A}}} {H_{0}^{\mathrm{A}}}
\,,
\end{equation}
such that $\delta(t) \ll 1$. Equation (\ref{Hsolution2}) thus
modifies to:
\begin{equation}\label{Hsolution3}
\Omega\left(t - t' \right) = - A \left[ \log \left(\frac{2 +
\delta(t)}{\left| \delta(t) \right| } \right) - c_{1} \right] + B
\left[ \log \left(\frac{ 1+ \delta(t) + H_{0}^{\mathrm{C}}/
H_{0}^{\mathrm{A}} }{\left| 1+ \delta(t) - H_{0}^{\mathrm{C}}/
H_{0}^{\mathrm{A}} \right|}\right)-c_{2} \right] \,,
\end{equation}
We can expand the second logarithm making use of $\lambda \ll 1$.
The leading order contribution (in $\delta(t)$) is given by the
denominator in the logarithm, because this term diverges as
$\delta(t)$ approaches zero. We can thus exponentiate the equation
and solve for the Hubble parameter:
\begin{equation}\label{Hsolution4}
\left| H_{>}(t) - H_{0}^{\mathrm{A}} \right| = 2
H_{0}^{\mathrm{A}} \exp\left[ \frac{\Omega}{A}\left(t - t' +\Delta
t_{>} \right) \right] \,,
\end{equation}
where the time shift $\Delta t_{>}$ is given by:
\begin{equation}\label{TimeShift1}
\Omega \Delta t_{>} = B \left( c_{2} - 2\frac{H_{0}^{\mathrm{C}}}
{H_{0}^{\mathrm{A}}} \right) - A c_{1}\,.
\end{equation}
As $\Omega >0$, this solution converges whenever $A<0$. This
provides a stability condition on $b''$ in terms of $b'$ and
$\lambda$. The quantum anomaly driven attractor is stable,
whenever the following inequality is satisfied:
\begin{equation}\label{StabilityQdSCond}
b'' > \frac{2}{9}b' \frac{4+3 \omega+16\pi\lambda b' (5+3\omega)}{
(1+\omega)(1+16\pi\lambda b') }\,.
\end{equation}
In figure \ref{fig:QdSInstability}, we have numerically calculated
the evolution of the Hubble parameter in a radiation dominated
Universe when this inequality is not satisfied. We used $b''=7b'/6
< 5 b'/6$. For initial conditions above the attractor, the Hubble
parameter increases to even higher energies, whereas for initial
conditions below the quantum attractor, the Hubble parameter
evolves towards the classical attractor. Hence even in quasi de
Sitter spacetimes, physically questionable solutions occur.
However, note that when $b''= 2b'/3$, the specific case under
consideration in subsection \ref{Case II: b''=2b'/3}, the above
inequality is satisfied.

When the above inequality is satisfied, the Hubble parameter in
the high energy limit decays exponentially towards the quantum
anomaly driven attractor, where some ''frequency dependence''
through $\Omega/A$ and a time shift $\Delta t_{>}$ can be
recognised. The time shift can without observational consequences
be absorbed in the initial time $t'$.

\begin{figure}
    \begin{minipage}[t]{.45\textwidth}
        \begin{center}
\includegraphics[width=\textwidth]{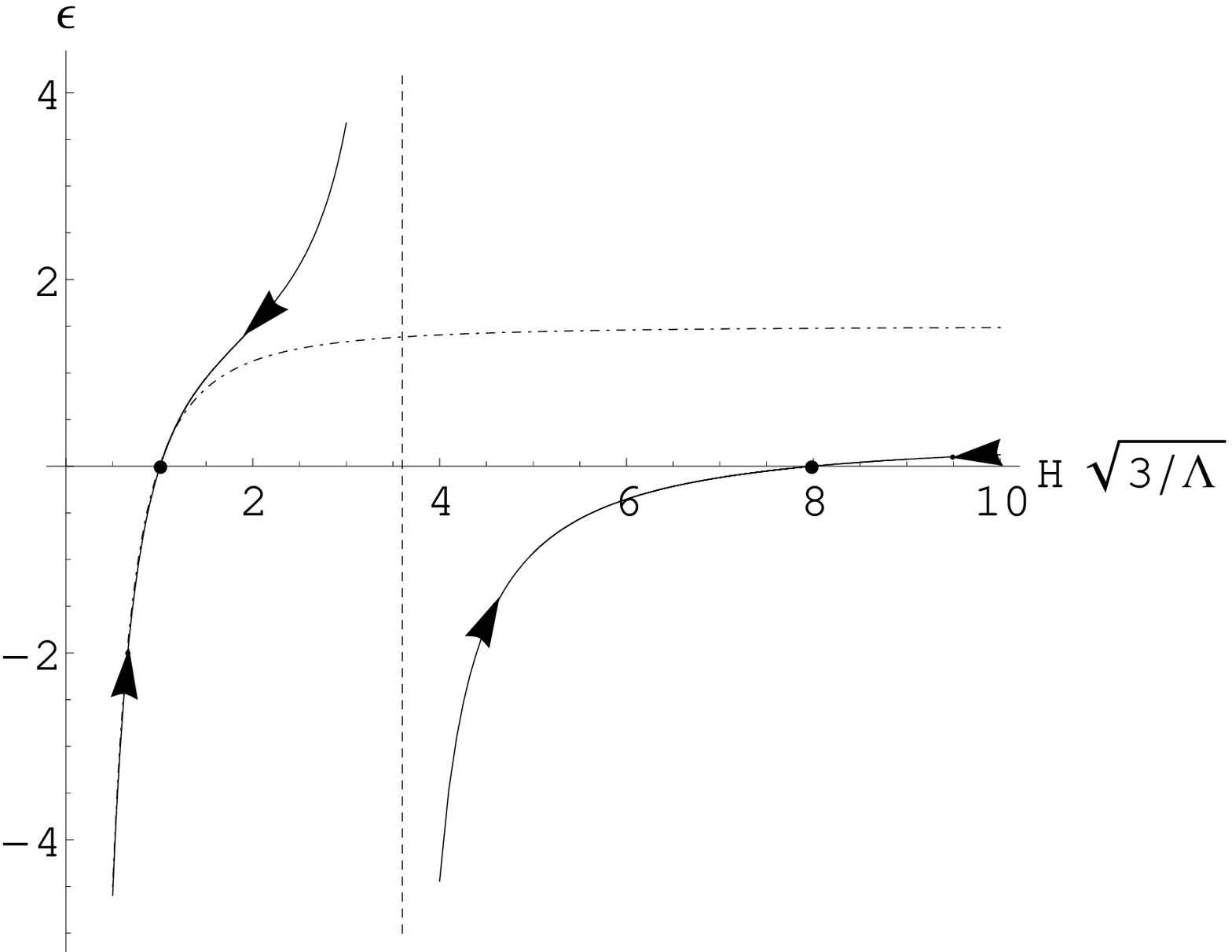}
   {\em \caption{Phase space flow in quasi de Sitter spacetime. The phase space
   consists of two lines. The flow, indicated by the arrows, is towards the classical
   or quantum attractor represented by two dots. In this regime, both attractors
   are stable. The vertical dashed line indicates the
   branching point and the dashed-dotted line the classical evolution.
   We have used $\omega=0$, $\lambda=1/50$, $b'=-0.015$ (Standard Model
   value) and $b''=0$. \label{fig:parametricHepsilon} }}
        \end{center}
    \end{minipage}
\hfill
    \begin{minipage}[t]{.45\textwidth}
        \begin{center}
\includegraphics[width=\textwidth]{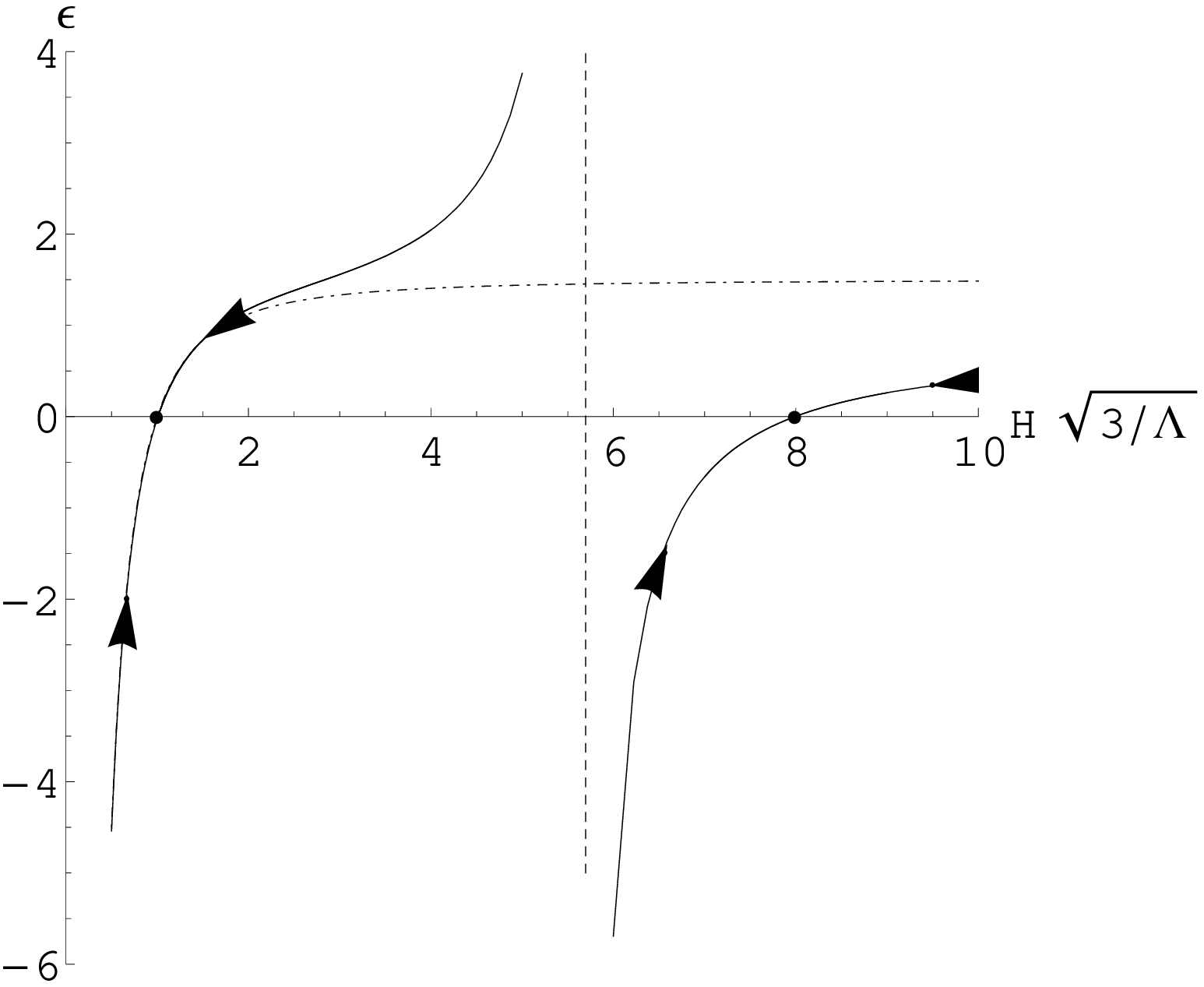}
   {\em \caption{Phase space flow in quasi de Sitter spacetime for $b''=2 b'/3$
   such that the $\Box R$ does not contribute in the trace anomaly. Qualitatively,
   the dynamics does not change when compared to figure \ref{fig:parametricHepsilon}.
   Again, both attractors are stable. Apart from $b''$, the value of the parameters are
   identical to figure \ref{fig:parametricHepsilon}.
   \label{fig:parametricHepsilonnoBoxR} }}
        \end{center}
    \end{minipage}
\end{figure}

The low energy limit reveals less surprising behaviour. Here, we
set:
\begin{equation}\label{Lowlimit}
\tilde{\delta}(t) = \frac{H(t) - H_{0}^{\mathrm{C}}} {
H_{0}^{\mathrm{C}}} \,,
\end{equation}
and $\tilde{\delta}(t) \ll 1$. Again we use $\lambda \ll 1$ in
order to capture the leading order dynamics. This yields:
\begin{equation}\label{Hsolution5}
\left| H_{<}(t) -  H_{0}^{\mathrm{C}} \right| = 2
H_{0}^{\mathrm{C}} \exp\left[-\frac{\Omega}{B}\left(t - t' +\Delta
t_{<}\right) \right] \,.
\end{equation}
Because $B>0$, this solution converges. In the low energy limit
the time shift $\Delta t_{<}$ is slightly different as compared to
(\ref{TimeShift1}):
\begin{equation}\label{TimeShift2}
\Omega \Delta t_{<} = B c_{2} - A \left( c_{1} -
2\frac{H_{0}^{\mathrm{C}}} {H_{0}^{\mathrm{A}}} \right) \,.
\end{equation}
Finally, we examine the phase space flow in quasi de Sitter
spacetime. In figure \ref{fig:parametricHepsilon}, we show a
parametric plot of $H$ versus $\epsilon$. The phase space
basically consists of two lines. The phase space flow is towards
either the classical or the quantum anomaly driven attractor as
indicated by the arrows. Note that we have chosen both attractors
to be stable. Furthermore, we also include the branching point
(\ref{BranchingPoint}) and the classical evolution, that is, the
evolution of a Universe with $b''=b'=0$. As expected, the flow is
towards the classical attractor in this case. Although the
analysis performed above is for generic values of $b''$, we set it
to zero in figure \ref{fig:parametricHepsilon} and $b'$ takes its
Standard Model value.

\subsection{Case II: $\mathbf{b''}=2 \mathbf{b'}/3$}
\label{Case II: b''=2b'/3}

As indicated earlier, we must consider the case when $b''=2 b'/3$
separately because in this particular case the total coefficient
in front of the $\Box R$ contribution to the trace anomaly
vanishes. All higher derivative contributions precisely cancel and
also the $\dot{H}^{2}$ contribution happens to cancel, such that
we find ourselves immediately situated in quasi de Sitter
spacetime. Albeit a simple case, we do take the full trace anomaly
into account.

The analytic solution obtained in (\ref{Hsolution1}) still
applies and moreover, it becomes exact. The branching point is
still given by equation (\ref{BranchingPoint}) for which we just
have to insert $b''=2b'/3$.
Clearly, in figure \ref{fig:parametricHepsilonnoBoxR} one
can see that qualitatively the dynamics has not changed compared
to figure \ref{fig:parametricHepsilon}. The branching point has
shifted somewhat to the right, and the way in which the Hubble
parameter approaches its two late time asymptotes differs.
However, the important features of figure
\ref{fig:parametricHepsilon}, i.e.: two stable attractors, the
occurrence of a branching point and the shape and dimension of the phase
space, do not change.

\section{The Trace Anomaly in FLRW Spacetimes}
\label{The Trace Anomaly in FLRW Spacetimes}

\begin{figure}[!ht]
    \begin{minipage}[t]{.45\textwidth}
        \begin{center}
\includegraphics[width=\textwidth]{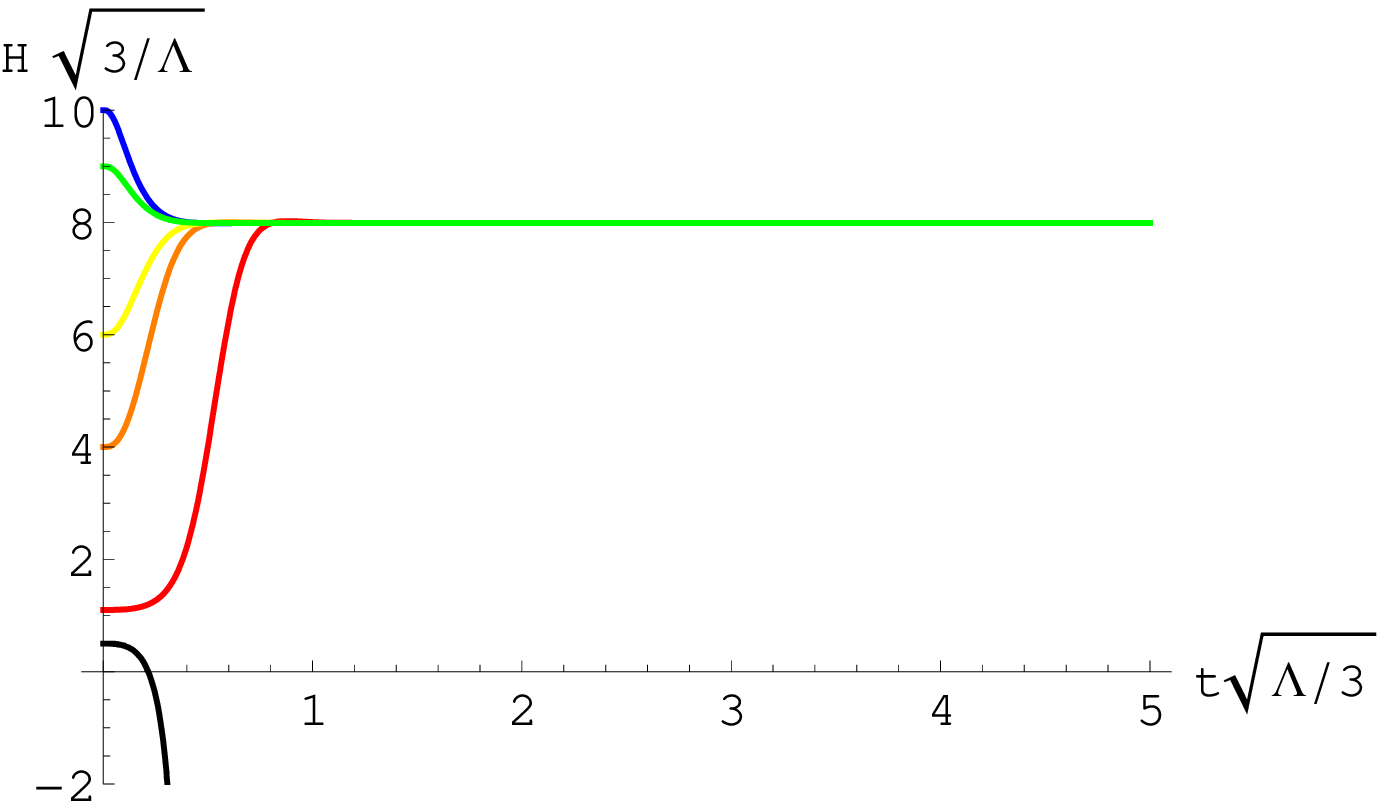}
   {\em \caption{Dynamics of the Hubble parameter taking the full trace anomaly
   into account. We took $b''=0$ such that $b''-2b'/3 > 0$
   yielding an unstable classical attractor. We have used $\omega=0$,
   $\lambda=1/50$, $b'=-0.015$ (Standard Model
   value). \label{fig:hubbleFullTA1} }}
\vskip 1.0cm
\includegraphics[width=\textwidth]{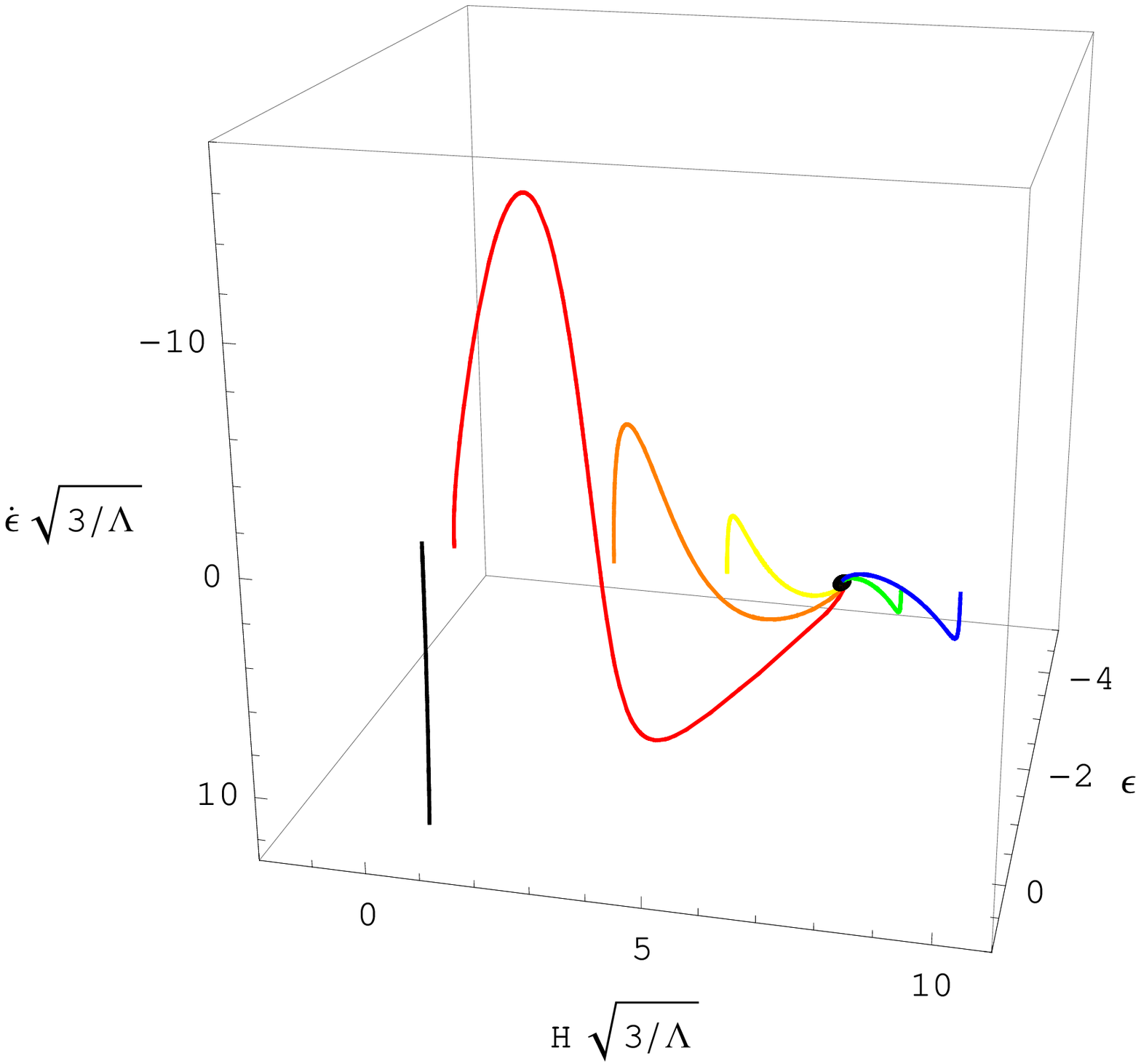}
   {\em \caption{Parametric phase space plot for figure \ref{fig:hubbleFullTA1} in which
   the classical attractor is unstable. We have indicated the quantum anomaly driven
   attractor as a small black sphere.
   \label{fig:phasespaceFullTA1} }}
        \end{center}
    \end{minipage}
\hfill
    \begin{minipage}[t]{.45\textwidth}
        \begin{center}
\includegraphics[width=\textwidth]{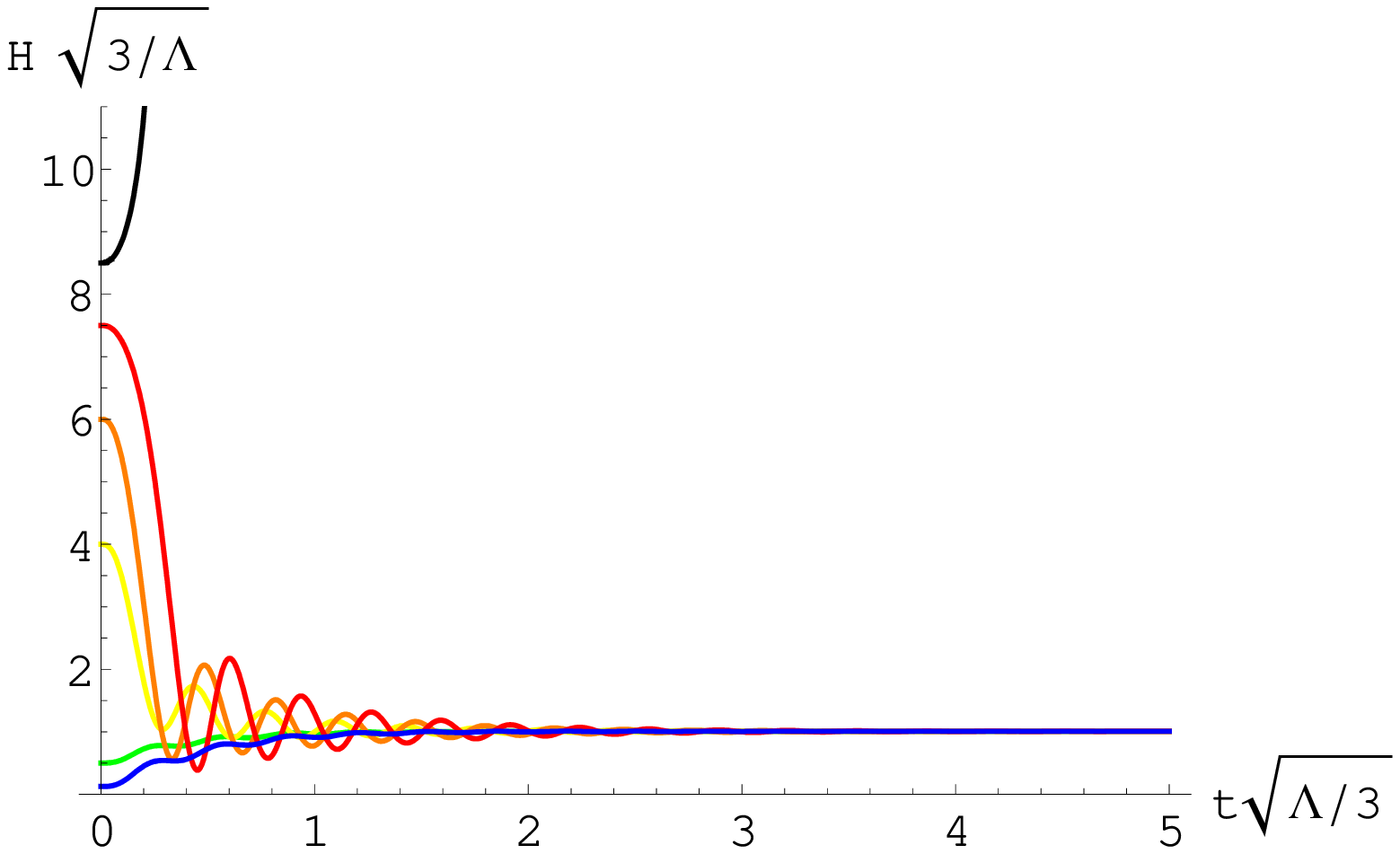}
   {\em \caption{Dynamics of the Hubble parameter taking the full trace anomaly
   into account. We took $b''=b'$ such that $b''-2b'/3 < 0$
   yielding a stable classical attractor. We have used $\omega=0$,
   $\lambda=1/50$, $b'=-0.015$ (Standard Model
   value). Clearly, the classical attractor is under-damped, resulting in various
   oscillations around $H_{0}^{\mathrm{C}}$.
   \label{fig:hubbleFullTA2} }}
\vskip 0.1cm
\includegraphics[width=\textwidth]{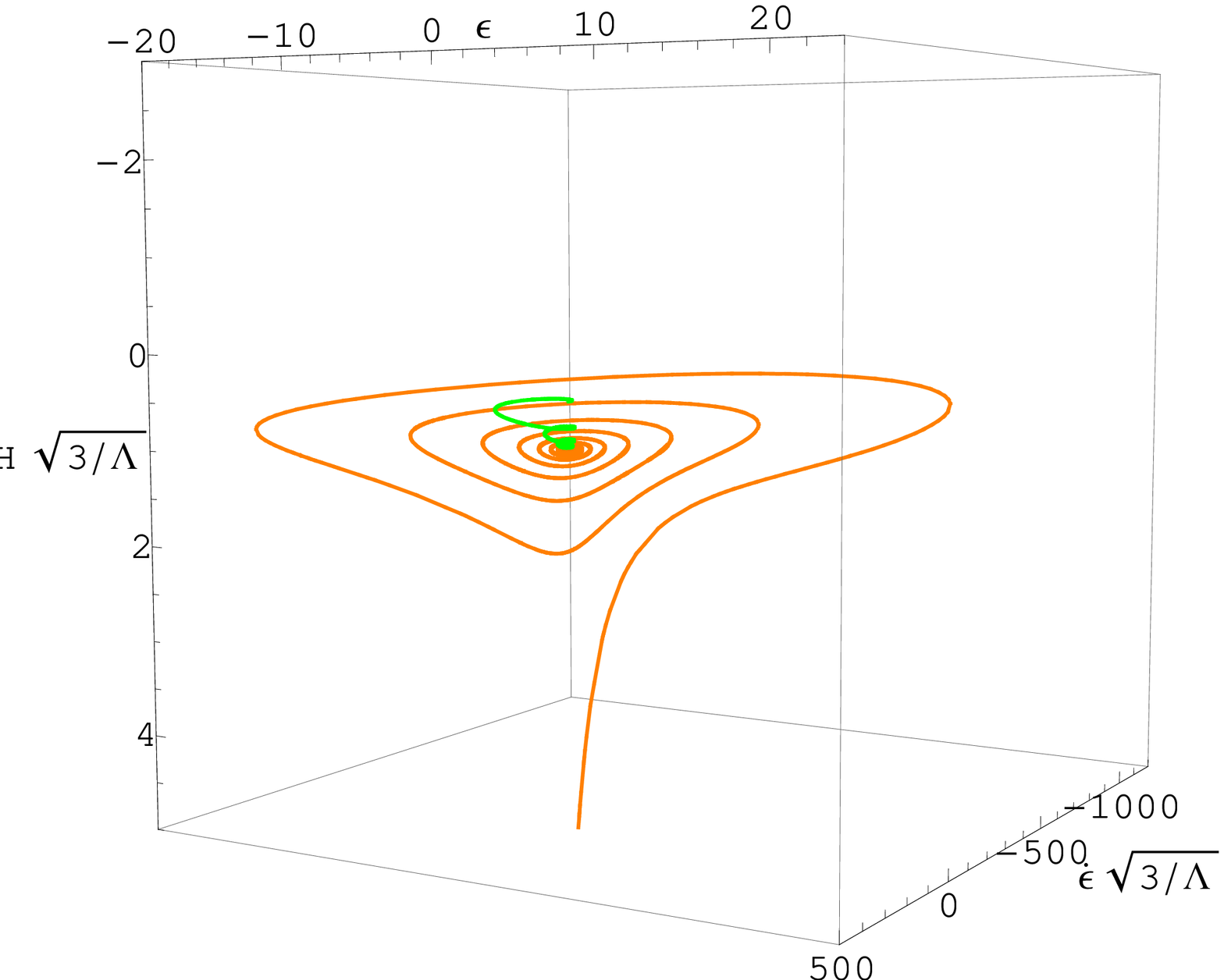}
   {\em \caption{Parametric phase space plot for figure \ref{fig:hubbleFullTA2} in which
   the classical attractor is stable. Because of under-damping,
   one spirals towards the classical attractor. For clarity, we
   have only included the flow for two initial conditions $H(0)=\sqrt{3/ \Lambda}/2$
   and $H(0)=6\sqrt{3/ \Lambda}$. The twister like structure is clearly visible.
   Qualitatively, the phase space flow resulting from other initial
   conditions is identical.
   \label{fig:phasespaceFullTA2} }}
        \end{center}
    \end{minipage}
\end{figure}

We turn our attention to solving the full trace equation
(\ref{EFEGeneral}), where we truncate the expression neither for
the anomalous trace (\ref{TraceAnomaly4}) nor for the quantum
density (\ref{quantumdensity3}). Obviously, we cannot solve this
equation analytically, for it contains all higher order derivative
contributions, which forces us to rely on numerical methods.

Firstly, we note that the asymptotes (\ref{Asymptote2}) do not
change by including the higher derivative contributions. Keeping
Ostrogradsky's theorem in mind, we expect to incur all kinds of
issues related to the stability of our system and asymptotes in
particular. It is therefore essential to perform a stability
analysis for small perturbations $\delta H(t)$ around both of the
asymptotes. We insert:
\begin{equation}\label{PerturbationAsymptote}
H(t)= H_{0}^{\mathrm{C,A}} + \delta H(t)\,,
\end{equation}
in equation (\ref{EFEGeneral}), where $H_{0}^{\mathrm{C,A}}$ can
either denote the classical or the quantum attractor. For the
small perturbations around these attractors we make the ansatz
$\delta H(t)=c\exp[\xi t]$. We linearise the trace equation
finding the characteristic equation from which we determine the
eigenvalues $\xi$ of our system:
\begin{eqnarray}\label{CharacteristicEquation}
2\left(\mu-3\nu/2 \right) \xi^{3} + 6 \left(\mu-3\nu/2
\right)(2+\omega) H_{0}^{\mathrm{C,A}} \xi^{2}+ \left(\left\{
6\mu(5+3\omega)-27\nu(1+\omega)\right\}
\left(H_{0}^{\mathrm{C,A}}\right)^2 -3\right)\xi && \nonumber\\
+ 9\left(4\mu
\left(H_{0}^{\mathrm{C,A}}\right)^2-1\right)(1+\omega)
H_{0}^{\mathrm{C,A}} &=&0 \,,
\end{eqnarray}
where $ \mu = -8\pi G b'$ and $ \nu = -8\pi G b''$. Remarkably,
the solutions of this third order equation are simple:
\begin{subequations}
\label{EigenvaluesStability}
\begin{eqnarray}
\xi^{(1)} &=& -3 H_{0}^{\mathrm{C,A}} (1+\omega)
\label{EigenvaluesStabilitya}\\
\xi^{(2)} &=& -\frac{3 H_{0}^{\mathrm{C,A}}}{2} + \sqrt{\Delta}
\label{EigenvaluesStabilityb}\\
\xi^{(3)} &=&  -\frac{3 H_{0}^{\mathrm{C,A}}}{2} - \sqrt{\Delta}
\label{EigenvaluesStabilityc}\,,
\end{eqnarray}
\end{subequations}
where:
\begin{equation}\label{DeltaDef}
\Delta= -\frac{3}{4 \left( 2\mu-3\nu\right) } \left\{
-4+\left(H_{0}^{\mathrm{C,A}}\right)^2 \left(10\mu+9\nu\right)
\right\} \,.
\end{equation}
Clearly, when $\mathrm{Re}\,\xi^{(i)}<0$ for $i=1,2,3$, the
corresponding attractor is stable. Since we only consider
non-tachyonic matter, eigenvalue (\ref{EigenvaluesStabilitya}) is
negative. However, a finite period in which $w<-1$ is not excluded
(see e.g.: \cite{Onemli:2004mb,Kahya:2006hc}). If one were to
consider other equations of state than the simple linear one
$\rho_{\mathrm{M}}=\omega p_{\mathrm{M}}$, this statement might no
longer hold \cite{Nojiri:2005sx}. Hence only for $\xi^{(2)}$ when
$\Delta>0$, we could encounter a potential instability.
Surprisingly, the stability analysis does not depend on the
equation of state $\omega$ because $\omega$ enters only through
equation (\ref{EigenvaluesStabilitya}). The condition for
instability thus reads:
\begin{equation}\label{CondInstab}
\Delta > \left(\frac{3H_{0}^{\mathrm{C,A}}}{2}\right)^2\,.
\end{equation}
We can rewrite this equation to find:
\begin{equation}\label{CondInstab2}
4 + 8\pi G\left(H_{0}^{\mathrm{C,A}}\right)^2
\left(10b'+9b''\right) \gtrless 72\pi G\left( b''-2b'/3 \right)
\left(H_{0}^{\mathrm{C,A}}\right)^2 \,.
\end{equation}
In the expression above, we should read the inequality $>$ or $<$
whenever $b''-2b'/3 > 0$ or $b''-2b'/3 < 0$, respectively. We can
now insert either the classical or quantum asymptotes previously
derived in equation (\ref{Asymptote2}) and verify which of the two
above inequalities is satisfied. Upon inserting the expression for
the classical attractor, equation (\ref{CondInstab2}) yields:
\begin{equation}\label{CondInstab3}
1 + 32 \pi \lambda b' \gtrless 0 \,,
\end{equation}
where $\lambda = G\Lambda/3 \ll 1$ as before. Only the first
inequality $>$ will be satisfied. Hence we conclude that the
classical attractor is unstable if $b''-2b'/3 > 0$. The converse
will be true if $b''-2b'/3 < 0$. Likewise, equation
(\ref{CondInstab2}) for the quantum attractor after some algebra
reads:
\begin{equation}\label{CondInstab4}
-1 - \frac{32 \pi \lambda}{3} b' \gtrless 0 \,,
\end{equation}
Concluding, when using $\lambda\ll 1$, we unambiguously find:
\begin{subequations}
\begin{eqnarray}
\mathrm{If} \,\, b''-2b'/3 > 0, && \mathrm{then}\,\, \left\{
\begin{array}{l}
\mbox{Classical attractor unstable} \\
\mbox{Quantum attractor stable}
\end{array} \right.
\label{StabilityCondition1}\\
\mathrm{If} \,\, b''-2b'/3 < 0, && \mathrm{then}\,\, \left\{
\begin{array}{l}
\mbox{Classical attractor stable} \\
\mbox{Quantum attractor unstable}
\end{array} \right.
\label{StabilityCondition2}
\end{eqnarray}
\end{subequations}
Let us first of all recall that it is precisely the combination
$b''-2b'/3$ that multiplies the $\Box R$ contribution in the trace
anomaly. This calculation thus proves the statements about
stability made in e.g. \cite{Pelinson:2002ef} using the
Routh-Hurwitz method. Our proof is more general because we include
a constant but otherwise arbitrary equation of state parameter
$\omega>-1$. Moreover, while the Routh-Hurwitz method can only
guarantee stability of a solution (when certain determinants are
all strictly positive), it does not tell anything about
instability \cite{Pelinson:2002ef, MathWorldRHTheorem}.
Furthermore appreciate that the singular point in this analysis,
$b''-2b'/3=0$, or equivalently $2\mu/3-\nu=0$, immediately directs
us to the quasi de Sitter spacetime analysis performed in section
\ref{Case II: b''=2b'/3}, where all higher derivative
contributions precisely cancel, rendering both attractors stable.

Let us compare figures \ref{fig:hubbleFullTA1} and
\ref{fig:hubbleFullTA2}. In the former figure, we used $b''=0$
such that $b''-2b'/3 > 0$, yielding an unstable classical
attractor. However, if $H(0) \leq H_{0}^{\mathrm{C}}$ the quantum
anomaly driven asymptote is not an attractor and the Hubble
parameter runs away to negative infinity. In the latter figure, we
set $b''=b'$ such that $b''-2b'/3 < 0$ which gives us a stable
classical attractor. Likewise, for initial conditions $H(0) \geq
H_{0}^{\mathrm{A}}$ the de Sitter solution is not an attractor and
the Hubble parameter rapidly blows up to positive infinity.

In figure \ref{fig:hubbleFullTA2}, we can observe another
interesting phenomenon. In this case, the classical attractor is
under-damped, resulting in decaying oscillations around the de
Sitter attractor. In figure \ref{fig:hubbleFullTA1} these
oscillations are not always present. The eigenvalues
(\ref{EigenvaluesStability}) develop an imaginary contribution
resulting in oscillatory behaviour whenever:
\begin{equation}\label{CondOscill}
\Delta < 0 \,.
\end{equation}
We thus find:
\begin{equation}\label{CondOscill2}
4 + 8\pi G\left(H_{0}^{\mathrm{C,A}}\right)^2
\left(10b'+9b''\right) \lessgtr 0 \,.
\end{equation}
The inequality $<$ or $>$ holds whenever $b''-2b'/3 > 0$ or
$b''-2b'/3 < 0$ applies, respectively. Again, we verify which of
the two inequalities is actually satisfied. To study oscillations
around a stable classical attractor, we should take the $>$
inequality (there are no oscillations around an unstable
attractor). We thus find:
\begin{equation}\label{CondOscill3}
1 + 2\pi \lambda \left(10b'+9b''\right) > 0 \,.
\end{equation}
Clearly, this inequality is always satisfied because $\lambda \ll
1$. We thus conclude that whenever the classical de Sitter
attractor is stable, oscillations occur. Furthermore, we can
insert the quantum anomaly driven attractor in equation
(\ref{CondOscill2}). Now, we should use the $<$ inequality in
order to study oscillatory behaviour around the (stable) quantum
attractor. This yields:
\begin{equation}\label{CondOscill4}
-1 - \frac{9}{2}\frac{b''}{b'} - 8 \pi \lambda
\left(10b'+9b''\right) < 0 \,.
\end{equation}
Oscillatory behaviour around the quantum attractor thus occurs
when:
\begin{equation}\label{CondOscill5}
b'' < - \frac{2}{9} b'\left( \frac{1+8\pi \lambda b'}{1 + 8 \pi
\lambda}\right)\,.
\end{equation}
\begin{figure}
    \begin{minipage}[t]{.5\textwidth}
        \begin{center}
\includegraphics[width=\textwidth]{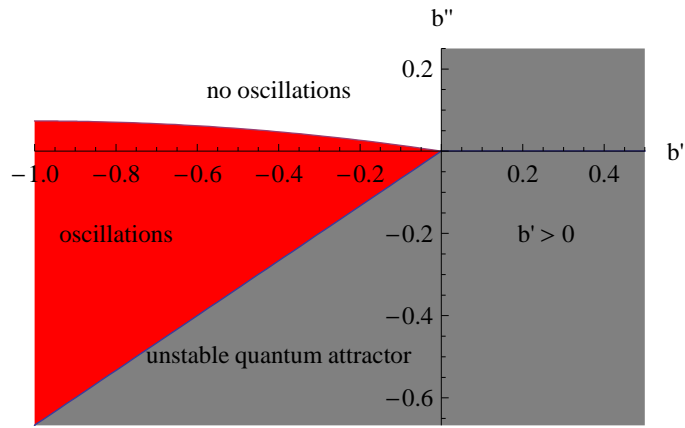}
   {\em \caption{Parameter space for oscillations around the
   quantum anomaly driven attractor. The gray region (light shaded) is excluded
   either because $b'>0$ or because the quantum attractor has become
   unstable. The red region (dark shaded) in parameter space shows oscillatory
   behaviour, whereas the white region is either critically damped
   or over-damped. We used $\lambda=1/50$. \label{fig:ParmSpaceOscil}}}
        \end{center}
    \end{minipage}
\end{figure}
In figure \ref{fig:ParmSpaceOscil} we depicted the parameter space
(the $b'$ versus $b''$ plane) resulting in oscillatory behaviour
around the quantum anomaly driven attractor. First of all, we
should have $b''-2b'/3 > 0$ yielding a stable quantum attractor.
Of course, one cannot have oscillations around an unstable
attractor. Secondly, note that $b'<0$ because of equation
(\ref{TraceAnomaly3b}). Finally, when the newly derived inequality
(\ref{CondOscill5}) is satisfied, oscillations occur. These
considerations divide the phase space into three regions: a region
where oscillatory behaviour occurs, another region which results
in critically or over-damped behaviour and a part of phase space
that is forbidden, as shown in figure \ref{fig:ParmSpaceOscil}.

Let us now return to discussing the classical attractor that shows
its oscillatory behaviour manifestly. We calculate the frequency
of oscillations around this attractor. Taking the square root of
(\ref{DeltaDef}) and extracting an $i$ we find the (quantum
corrected) frequency for oscillations around the classical
attractor:
\begin{equation}\label{FreqCLOscil}
\omega_{\mathrm{C}} = \sqrt{\frac{1+2 \pi\lambda
\left(10b'+9b''\right)} {8\pi G\left(2b'/3 - b'' \right)}} \,.
\end{equation}
Note that this frequency is independent on the equation of state
parameter $\omega$. The frequency of oscillations around the
quantum attractor can be found analogously.

Let us analyse the phase space in the case of a stable quantum and
classical attractor subsequently. In figure
\ref{fig:phasespaceFullTA1}, we visualise the phase space flow for
the former case parametrically in the $H(t)$, $\epsilon(t)$ and
$\dot{\epsilon}(t)$ directions\footnote{Note we are not able to
include the fourth dimension of the phase space,
$\ddot{\epsilon}(t)$. However, also $\ddot{\epsilon}(t)$ rapidly
approaches zero as time elapses.}. The small black sphere denotes
the quantum anomaly driven attractor. For the latter case, we
include in figure \ref{fig:phasespaceFullTA2} the phase space flow
for just two initial conditions for $H(t)$ for clarity. The
under-damped oscillatory behaviour results in the twister like
structure visible in the $H(t)$, $\epsilon(t)$ and
$\dot{\epsilon}(t)$ directions.

Finally, we would like to point out that including all higher
derivative contributions, which thus corresponds to solving the
trace equation exactly, modifies the dynamics of the Hubble
parameter significantly. Attractors that were stable in the
absence of higher derivatives under certain conditions
destabilise. The reason for this, clearly, is attributable to the
presence of the $\Box R$ term in the trace anomaly, generating
these higher derivative contributions. We do not know whether or
not incorporating the higher derivatives is a sensible thing to
do. Usually higher order derivatives tend to destabilise a system
signifying that some particular solutions are not physical.
Therefore, in the spirit of Ostrogradsky's theorem, one can
question whether the analysis where higher derivative
contributions are discarded is correct, or the analysis taking the
full trace anomaly into account.

\section{Conclusion}
\label{Conclusion}

The trace of the Einstein field equations in cosmologically
relevant spacetimes together with stress-energy conservation
completely captures the dynamics of the Hubble parameter. We have
derived the trace anomaly from an effective action in spatially
flat FLRW spacetimes. It consists of the local quadratic geometric
curvature invariants $R^2$ and the Gauss-Bonnet term $E$. Because
of counterterms that are supposed to cancel divergences of the as
yet unknown underlying fundamental theory, we expect the
coefficients in the trace anomaly to change. The physical value of
each of these coefficients receives contributions both from the
anomalous trace and from these counterterms. Because we do not
know the physical value these parameters will take, we must allow
them to vary in order to examine all possibilities.

We have studied the dynamics of the Hubble parameter both in quasi
de Sitter and in FLRW spacetimes including matter, a cosmological
term and the trace anomaly. In quasi de Sitter spacetime, where we
restrict the Hubble parameter to vary slowly in time, we find that
for various initial conditions $H(t)$ asymptotes either to the
classical de Sitter attractor, or to a quantum anomaly driven
attractor. We find a region in parameter space where the quantum
attractor destabilises. Otherwise, both attractors are stable.

In FLRW spacetimes we include all higher derivative contributions
in the trace anomaly. We perform a stability analysis for small
perturbations around the two asymptotes. For $b''-2b'/3
> 0$, the quantum attractor is stable and the classical de Sitter
attractor is unstable. On the contrary, for $b''-2b'/3 < 0$, the
quantum attractor is unstable and the de Sitter attractor becomes
stable. The singular point in this analysis, $b''-2b'/3 = 0$,
immediately directs us to quasi de Sitter spacetime in which the
dynamics is much simpler. In this case, both attractors are
stable. The classical de Sitter attractor always shows
under-damped oscillatory behaviour and we calculate the frequency
of these oscillations. We analyse the phase space of the quantum
attractor and conclude there is some region in parameter space for
which oscillations occur.

There is no dynamical effect that influences the effective value
of the cosmological constant, i.e.: the classical de Sitter
attractor. Based on our semiclassical analysis we thus conclude
that the trace anomaly does \textit{not} solve the cosmological
constant problem.

We have studied both the truncated and the exact expression of the
trace anomaly in flat FLRW spacetimes. We do not know which of the
two approaches is correct. Keeping Ostrogradsky's theorem in mind,
higher derivative contributions usually have the tendency to
destabilise a dynamical system. Discarding these higher
derivatives and studying the trace anomaly in quasi de Sitter
spacetime would thus seem plausible.

Finally, one could wonder whether the quantum anomaly driven
attractor is physical. The quantum attractor is of the order of
the Planck mass $M_{\mathrm{pl}}$, so only when matter in the
early Universe is sufficiently dense, $H \simeq \mathcal{O}(
M_{\mathrm{pl}})$. We then expect to evolve towards the quantum
attractor. However, at these early times we also expect
perturbative general relativity to break down. Hence, this
attractor might even not be there or it may be seriously affected
by quantum fluctuations. Quantum fluctuations present at that
epoch might even induce tunnelling towards the regime where $H(t)$
asymptotes to the classical attractor.

\

\noindent \textbf{Acknowledgements}

\noindent We thank Tomas Janssen and Jan Smit for useful
suggestions. The authors acknowledge financial support from FOM
grant 07PR2522 and by Utrecht University.


\begin{thebibliography}{99}
\bibitem{Nobbenhuis:2004wn}
  S.~Nobbenhuis,
  Categorizing Different Approaches to the Cosmological Constant Problem,
  Found.\ Phys.\  {\bf 36} (2006) 613
  [arXiv:gr-qc/0411093].
\bibitem{Nobbenhuis:2006yf}
  S.~Nobbenhuis,
  The Cosmological Constant Problem, an Inspiration for New Physics,
  Ph.D. Thesis,
  [arXiv:gr-qc/0609011].
\bibitem{Bousso:2007gp}
  R.~Bousso,
  TASI Lectures on the Cosmological Constant,
  [arXiv:hep-th/0708.4231].
\bibitem{Donoghue:1996ma}
  J.~Donoghue,
  The Quantum Theory of General Relativity at Low Energies,
  Helv.\ Phys.\ Acta {\bf 69} (1996) 269
  [arXiv:gr-qc/9607039].
\bibitem{Donoghue:1997hx}
  J.~Donoghue,
  Perturbative Dynamics of Quantum General Relativity,
  [arXiv:gr-qc/9712070].
\bibitem{Birrell:1982ix}
  N.~D.~Birrell and P.~C.~W.~Davies,
  Quantum Fields in Curved Space,
  Cambridge monographs on Mathematical Physics,
  Cambridge University Press (1982)
\bibitem{Capper:1974ic}
  D.~M.~Capper and M.~J.~Duff,
  Trace Anomalies in Dimensional Regularization,
  Nuovo Cim.\  A {\bf 23}, 173 (1974).
\bibitem{Deser:1976yx}
  S.~Deser, M.~J.~Duff and C.~J.~Isham,
  Nonlocal Conformal Anomalies,
  Nucl.\ Phys.\  B {\bf 111}, 45 (1976).
\bibitem{Brown:1976wc}
  L.~S.~Brown,
  Stress Tensor Trace Anomaly in a Gravitational Metric: Scalar Fields,
  Phys.\ Rev.\  D {\bf 15} (1977) 1469.
\bibitem{Dowker:1976zf}
  J.~S.~Dowker and R.~Critchley,
  The Stress Tensor Conformal Anomaly for Scalar and Spinor Fields,
  Phys.\ Rev.\  D {\bf 16} (1977) 3390.
\bibitem{Tsao:1977tj}
  H.~S.~Tsao,
  Conformal Anomalies in a General Background Metric,
  Phys.\ Lett.\  B {\bf 68} (1977) 79.
\bibitem{Duff:1977ay}
  M.~J.~Duff,
  Observations on Conformal Anomalies,
  Nucl.\ Phys.\  B {\bf 125}, 334 (1977).
\bibitem{Brown:1977pq}
  L.~S.~Brown and J.~P.~Cassidy,
  Stress Tensor Trace Anomaly in a Gravitational Metric: General Theory,
  Maxwell Field,
  Phys.\ Rev.\  D {\bf 15} (1977) 2810.
\bibitem{Duff:1993wm}
  M.~J.~Duff,
  Twenty Years of the Weyl Anomaly,
  Class.\ Quant.\ Grav.\  {\bf 11}, 1387 (1994)
  [arXiv:hep-th/9308075].
\bibitem{Ford:1984hs}
  L.~H.~Ford,
  Quantum Instability of de Sitter Space-Time,
  Phys.\ Rev.\  D {\bf 31} (1985) 710.
\bibitem{Antoniadis:1986sb}
  I.~Antoniadis and E.~Mottola,
  Graviton Fluctuations in de Sitter Space,
  J.\ Math.\ Phys.\  {\bf 32} (1991) 1037.
\bibitem{Tsamis:1992xa}
  N.~C.~Tsamis and R.~P.~Woodard,
  The Structure of Perturbative Quantum Gravity on a de Sitter Background,
  Commun.\ Math.\ Phys.\  {\bf 162} (1994) 217.
\bibitem{Tsamis:1996qq}
  N.~C.~Tsamis and R.~P.~Woodard,
  Quantum Gravity Slows Inflation,
  Nucl.\ Phys.\  B {\bf 474} (1996) 235
  [arXiv:hep-ph/9602315].
\bibitem{Tsamis:1996qk}
  N.~C.~Tsamis and R.~P.~Woodard,
  One Loop Graviton Self-Energy in a Locally de Sitter Background,
  Phys.\ Rev.\  D {\bf 54} (1996) 2621
  [arXiv:hep-ph/9602317].
\bibitem{Abramo:1997hu}
  L.~R.~W.~Abramo, R.~H.~Brandenberger and V.~F.~Mukhanov,
  The Energy-momentum Tensor for Cosmological Perturbations,
  Phys.\ Rev.\  D {\bf 56} (1997) 3248
  [arXiv:gr-qc/9704037].
\bibitem{Finelli:2004bm}
  F.~Finelli, G.~Marozzi, G.~P.~Vacca and G.~Venturi,
  Adiabatic Regularization of the Graviton Stress-energy Tensor in de Sitter
  space-time,
  Phys.\ Rev.\  D {\bf 71} (2005) 023522
  [arXiv:gr-qc/0407101].
\bibitem{Bilandzic:2007nb}
  A.~Bilandzic and T.~Prokopec,
  Quantum Radiative Corrections to Slow-roll Inflation,
  Phys.\ Rev.\  D {\bf 76} (2007) 103507
  [arXiv:0704.1905 [astro-ph]].
\bibitem{Janssen:2007ht}
  T.~Janssen and T.~Prokopec,
  A Graviton Propagator for Inflation,
  Class.\ Quant.\ Grav.\ {\bf 25} (2008) 055007
  [arXiv:0707.3919 [gr-qc]].
\bibitem{Janssen:2008dw}
  T.~Janssen and T.~Prokopec,
  Implications of the Graviton One-loop Effective Action on the Dynamics of
  the Universe,
  arXiv:0807.0447 [gr-qc].
\bibitem{Janssen:2008dp}
  T.~Janssen, S.~P.~Miao and T.~Prokopec,
  Graviton One-loop Effective Action and Inflationary Dynamics,
  arXiv:0807.0439 [gr-qc].
\bibitem{Bilic:2007gr}
  N.~Bilic, B.~Guberina, R.~Horvat, H.~Nikolic and H.~Stefancic,
  On Cosmological Implications of Gravitational Trace Anomaly,
  Phys.\ Lett.\  B {\bf 657} (2007) 232
  [arXiv:gr-qc/0707.3830].
\bibitem{Schutzhold:2002pr}
  R.~Schutzhold,
  Small Cosmological Constant from the QCD Trace Anomaly?,
  Phys.\ Rev.\ Lett.\  {\bf 89} (2002) 081302.
\bibitem{Tomboulis:1988gw}
  E.~T.~Tomboulis,
  Dynamically Adjusted Cosmological Constant and Conformal Anomalies,
  Nucl.\ Phys.\  B {\bf 329} (1990) 410.
\bibitem{Antoniadis:1991fa}
  I.~Antoniadis and E.~Mottola,
  4-D Quantum Gravity in the Conformal Sector,
  Phys.\ Rev.\  D {\bf 45} (1992) 2013.
\bibitem{Antoniadis:1992hz}
  I.~Antoniadis,
  Dynamics of the Conformal Factor in 4-D Gravity,
  [arXiv:hep-th/9211055].
\bibitem{Antoniadis:1998fi}
  I.~Antoniadis, P.~O.~Mazur and E.~Mottola,
  Fractal Geometry of Quantum Spacetime at Large Scales,
  Phys.\ Lett.\  B {\bf 444} (1998) 284
  [arXiv:hep-th/9808070].
\bibitem{Salehi:2000eu}
  H.~Salehi and Y.~Bisabr,
  Conformal Anomaly and Large Scale Gravitational Coupling,
  Int.\ J.\ Theor.\ Phys.\  {\bf 39} (2000) 1241
  [arXiv:hep-th/0001095].
\bibitem{Antoniadis:2006wq}
  I.~Antoniadis, P.~O.~Mazur and E.~Mottola,
  Cosmological Dark Energy: Prospects for a Dynamical Theory,
  New J.\ Phys.\  {\bf 9} (2007) 11
  [arXiv:gr-qc/0612068].
\bibitem{Riegert:1984kt}
  R.~J.~Riegert,
  A Nonlocal Action for the Trace Anomaly,
  Phys.\ Lett.\  B {\bf 134}, 56 (1984).
\bibitem{Komatsu:2008hk}
  E.~Komatsu {\it et al.}  [WMAP Collaboration],
  Five-Year Wilkinson Microwave Anisotropy Probe (WMAP)
  Observations: Cosmological Interpretation,
  arXiv:0803.0547 [astro-ph].
\bibitem{Starobinsky:1980te}
  A.~A.~Starobinsky,
  A New Type of Isotropic Cosmological Models without Singularity,
  Phys.\ Lett.\  B {\bf 91} (1980) 99.
\bibitem{Hawking:2000bb}
  S.~W.~Hawking, T.~Hertog and H.~S.~Reall,
  Trace Anomaly Driven Inflation,
  Phys.\ Rev.\  D {\bf 63} (2001) 083504
  [arXiv:hep-th/0010232].
\bibitem{Fabris:2000mq}
  J.~C.~Fabris, A.~M.~Pelinson and I.~L.~Shapiro,
  Anomaly-induced Effective Action and Inflation,
  Nucl.\ Phys.\ Proc.\ Suppl.\  {\bf 95}, 78 (2001)
  [arXiv:hep-th/0011030].
\bibitem{Shapiro:2003gm}
  I.~L.~Shapiro,
  An Overview of the Anomaly-induced Inflation,
  Nucl.\ Phys.\ Proc.\ Suppl.\  {\bf 127} (2004) 196
  [arXiv:hep-ph/0311307].
\bibitem{Shapiro:2004wt}
  I.~L.~Shapiro,
  Effective Field Theory and Fundamental Interactions,
  [arXiv:hep-th/0412115].
\bibitem{Shapiro:2002nz}
  I.~L.~Shapiro and J.~Sola,
  A Modified Starobinsky's Model of Inflation: Anomaly-induced Inflation,
  SUSY and Graceful Exit,
  [arXiv:hep-ph/0210329].
\bibitem{Pelinson:2002ef}
  A.~M.~Pelinson, I.~L.~Shapiro and F.~I.~Takakura,
  On the Stability of the Anomaly-induced Inflation,
  Nucl.\ Phys.\  B {\bf 648}, 417 (2003)
  [arXiv:hep-ph/0208184].
\bibitem{Pelinson:2003gn}
  A.~M.~Pelinson, I.~L.~Shapiro and F.~I.~Takakura,
  Stability Issues in the Modified Starobinsky Model,
  Nucl.\ Phys.\ Proc.\ Suppl.\  {\bf 127}, 182 (2004)
  [arXiv:hep-ph/0311308].
\bibitem{Brevik:2006nh}
  I.~Brevik and J.~Quiroga Hurtado,
  Vanishing Cosmological Constant in Modified Gauss-Bonnet Gravity with
  Conformal Anomaly,
  Int.\ J.\ Mod.\ Phys.\  D {\bf 16} (2007) 817
  [arXiv:gr-qc/0610044].
\bibitem{Woodard:2006nt}
  R.~P.~Woodard,
  Avoiding Dark Energy with 1/R Modifications of Gravity,
  Lect.\ Notes Phys.\  {\bf 720} (2007) 403
  [arXiv:astro-ph/0601672].
\bibitem{Fischetti:1979ue}
  M.~V.~Fischetti, J.~B.~Hartle and B.~L.~Hu,
  Quantum Effects in the Early Universe. 1. Influence of Trace Anomalies on
  Homogeneous, Isotropic, Classical Geometries,
  Phys.\ Rev.\  D {\bf 20} (1979) 1757.
\bibitem{Onemli:2004mb}
  V.~K.~Onemli and R.~P.~Woodard,
  Quantum Effects can Render $w < -1$ on Cosmological Scales,
  Phys.\ Rev.\  D {\bf 70} (2004) 107301
  [arXiv:gr-qc/0406098].
\bibitem{Kahya:2006hc}
  E.~O.~Kahya and V.~K.~Onemli,
  Quantum Stability of a $w < - 1$ Phase of Cosmic Acceleration,
  Phys.\ Rev.\  D {\bf 76} (2007) 043512
  [arXiv:gr-qc/0612026].
\bibitem{Shapiro:2006sy}
  I.~L.~Shapiro,
  Local Conformal Symmetry and its Fate at Quantum Level,
  PoS {\bf IC2006} (2006) 030
  [arXiv:hep-th/0610168].
\bibitem{Shapiro:2008sf}
  I.~L.~Shapiro,
  Effective Action of Vacuum: Semiclassical Approach,
  Class.\ Quant.\ Grav.\  {\bf 25}, 103001 (2008)
  [arXiv:0801.0216 [gr-qc]].
\bibitem{Davies:1977ti}
  P.~C.~W.~Davies,
  Singularity Avoidance and Quantum Conformal Anomalies,
  Phys.\ Lett.\  B {\bf 68} (1977) 402.
\bibitem{Gorbar:2002pw}
  E.~V.~Gorbar and I.~L.~Shapiro,
  Renormalization Group and Decoupling in Curved Space,
  JHEP {\bf 0302} (2003) 021
  [arXiv:hep-ph/0210388].
\bibitem{Gorbar:2003yt}
  E.~V.~Gorbar and I.~L.~Shapiro,
  Renormalization Group and Decoupling in Curved Space. II: The Standard
  Model and Beyond,
  JHEP {\bf 0306} (2003) 004
  [arXiv:hep-ph/0303124].
\bibitem{Nojiri:2005sx}
  S.~Nojiri, S.~D.~Odintsov and S.~Tsujikawa,
  Properties of Singularities in (Phantom) Dark Energy Universe,
  Phys.\ Rev.\  D {\bf 71} (2005) 063004
  [arXiv:hep-th/0501025].
\bibitem{MathWorldRHTheorem}
  E.~W.~Weisstein, "Routh-Hurwitz Theorem.", From MathWorld -- A
  Wolfram Web Resource,\\
  {\tt http://mathworld.wolfram.com/Routh-HurwitzTheorem.html}.
\end{thebibliography}
\end{document}